# Elementary Steps of Energy Conversion in Strongly Correlated Systems: Beyond Single Quasiparticles and Rigid Bands

V. Moshnyaga[1], Ch. Jooss[2,7], P. E. Blöchl[3,4], V. Bruchmann-Bamberg[1], A. Dehning[2], L. Allen-Rump[3,4], C. Hausmann[2], M. Krüger[4], A. Rathnakaran[4], S. Rajpurohit[5], D. Steil[1], C. Flathmann[6], J. Hoffmann[2], M. Seibt[6], C. Volkert[2]

[1]University of Göttingen, 1st Institute of Physics – Friedrich-Hund-Platz 1, 37077 Göttingen, Germany

[2]University of Göttingen, Institute of Materials Physics, Friedrich-Hund-Platz 1, 37077 Göttingen, Germany

[3]Clausthal University of Technology, Institute of Theoretical Physics, 38678 Clausthal-Zellerfeld, Germany

[4]University of Göttingen, Institute of Theoretical Physics, Friedrich-Hund-Platz 1, 37077 Göttingen, Germany

[5]Molecular Foundry and Materials Sciences Division, Lawrence Berkeley National Laboratory, California 94720, USA

[6]University of Göttingen, 4th Institute of Physics – Solids and Nanostructures, Friedrich-Hund-Platz 1, 37077 Göttingen, Germany

[7]University of Göttingen, International Center for Advanced Studies of Energy Conversion (ICASEC), Göttingen D-37077, Germany

**Abstract**

Energy conversion in materials can be considered as a sequence of elementary steps initiated by a primary excitation. While these steps are quite well understood in classical semiconductors in terms of quasiparticle (QP) excitations and interactions, their understanding in strongly correlated materials is still elusive. Here, we review the progress which has been achieved over recent years by studies of manganite perovskite oxides as a model system for materials with strong correlations. They show a subtle interplay of different types of correlations, i.e., electron-phonon, electron-electron and spin-spin, resulting in rich physical phenomena due to competition between different ground states accompanied by temperature- and field-induced phase transitions. They strongly impact various types of energy conversion and transport processes including friction at surfaces, thermal transport, time-, energy- and power-dependent optical excitations as well as photovoltaic energy conversion. The underlying microscopic processes can be broken down to the behavior of the low-energy thermal and high-energy optical excitations, their interactions, transport and conversion which are theoretically analyzed by using models of interacting and tunable QPs: Their nature and interactions can change during excitation, transport and phase transitions, thus modifying electronic structure. At sufficiently high stimulation, QP excitations can even induce or actuate phase transitions. As a result, we obtained a comprehensive understanding of energy conversion steps going far beyond single QP pictures and rigid band approximations well-known for conventional semiconductors.

Table of Content

# 1. Introduction: Excitation, relaxation and transformation of interacting QPs

Energy conversion in many solids, i.e. excitation, relaxation, transport and transformation, as a response to an external stimulation has been successfully described by models, where the complex dynamics of a many body system of atoms and electrons can be decoupled and mapped to single particle excitations. The energy of their ground and excited states is described by rigid bands that are determined by the average equilibrium positions of the atoms[1],[2]. Such single respective independent particle concepts are powerful tools, taking into account the response of the medium to the particle, such as polarization and relaxation clouds of electrons and lattices in the concept of quasiparticles (QPs) as introduced by Landau[3],[4]. They can be described as dressed electrons or composite objects that move through the system in form of a Bloch wave in a dispersive conduction band. The same holds true when instead a hole is added to the system, i.e., when an electron is removed.

In materials with strong electronic correlations, however, finding such approximations is challenging since the interactions of QPs can give rise to different types of order. Moreover, QP excitations can result in a change of type of QPs and these effects can lead to qualitatively new behavior such as phase transitions driven by changing QPs interactions[5],[6]. Hence, the established models based on rigid band states populated by single particle excitations and excitation-independent interactions can break down in the correlated materials. Although strongly correlated systems are not yet integrated into applications, some of the mechanisms of energy conversion studied in a correlated system are also relevant for materials in near-future energy applications, such as halides[7],[8] or organic solar cells[9],[10]. Thus, we believe it is the right time and scope of this article to critically review the progress achieved in the fundamental understanding of elementary steps of energy conversion in strongly correlated systems.

Transition metal oxides (TMO) with perovskite structure, like cuprates, nickelates and manganites are among the most instructive and interesting model systems in this context. They exhibit a plethora of functionalities and phase transitions, arising from strong electronic correlations including electron-phonon, exchange and spin-phonon coupling. For example, cuprates ($YBa_2Cu_3O_{7-\delta}$) display high-$T_C$ superconductivity[11], nickelates ($NdNiO_3$) are known for coupled metal-insulator and structural phase transitions[12] and manganites (($La_{1-y}Pr_y)_{1-x}Ca_xMnO_3$) for the colossal magnetoresistance[13] (CMR) effect as well as for photo-induced phase transitions[14]. The properties of TMO materials, including their ground states and phase transitions, can be continuously varied and finely tuned by changing their chemical composition by doping and/or substitution[15],[16] as well as by applying external stimuli, like temperature, pressure/strain[17] or electromagnetic radiation[18]. Such versatility of ground states which only slightly differ in their free energy in combination with the controlling tools makes TMO especially interesting as model systems for experimental and theoretical studies of *excitations, transport* and *relaxations* of QPs, their *interactions* and *transformations*, which all can be viewed as elementary steps of energy conversion.

Central model systems in this review are perovskite manganites with general formula $RE_{1-x}A_xMnO_3$, where the A-sites of the perovskite structure are occupied by the rare earth RE=La, Pr, Nd, etc. and by alkali cations A=Sr, Ca, Ba. The different oxidation states of A-cations, i.e. $RE^{3+}$ and $A^{2+}$, and their different cation radii, e.g., $R_{La}>R_{Pr}$ within the mixed A-site occupation allow to carry out both hole doping for $0<x<0.5$ and chemical-pressure-driven bandwidth control, respectively. In addition, the dimensionality of the system and thus of the interactions can be tuned in the layered Ruddlesden-Popper (RP) variants $(RE, A)_{n+1}B_nO_{3n+1}$, with "n" representing the number of perovskite layers within the unit cell[19]. As a result, perovskite manganites display a rich variety of ground states with different magnetic (ferro-/antiferromagnetic, spin glass), structural (rhombohedral, orthorhombic, monoclinic) and electric (metal, insulator, charge/orbital ordering) phases summarized in different phase diagrams[20],[15a],[21],[22],[23] (see Fig. 1). A competition of the two fundamental tendencies in the correlation physics, i.e. localization and delocalization of charge carriers, as well as degeneracy of phases lying in close proximity within the corresponding phase diagram, leads to nanoscale electronic phase separation (EPS), which is a hallmark of perovskite manganites[24]. The field- and temperature-driven competition of different phases within the EPS regime results in the "colossal" magnetoresistance[13 above] and electroresistance[25] observed by applying magnetic and electric fields, respectively.

Another hallmark of manganites, discovered during a more than 50-year history of manganite research, is a new type of QP called Jahn-Teller (JT) polaron[26]. This QP originates from a specific type of electron-phonon coupling, i.e. the lifting of degeneracy of the d states of transition metal oxides coupled to a structural distortion of the oxygen ligand field[27]. It is an important ingredient of the electron-lattice correlations of many transition metal oxides[28], in particular of nickelates, manganites and cuprates[29]. In manganites it can be understood as an electron trapped by the JT distortion of the $MnO_6$ octahedron. The formation of trapped states is the basis for the ionic picture, where $Mn^{3+}$ is the JT ion and hole doping by substitution of La by Sr/Ca changes the concentration of JT ions. In the parent compound $LaMnO_3$ the whole perovskite structure is coherently JT distorted, showing orbital order[30]. Depending on the amount of JT ions, crystal structure (rhombohedral or orthorhombic) and strength of electron-phonon coupling the JT distortions can be static or dynamic[31]. The corresponding JT polarons can be single non-interacting polarons (the amount of JT ions is small like in an optimally doped rhombohedral LSMO, Fig. 1 a)) or correlated polarons with two or more QPs bound to each other at low enough temperature and within an orthorhombic structure like in LCMO (Fig. 1 b)) and LPCMO[32].

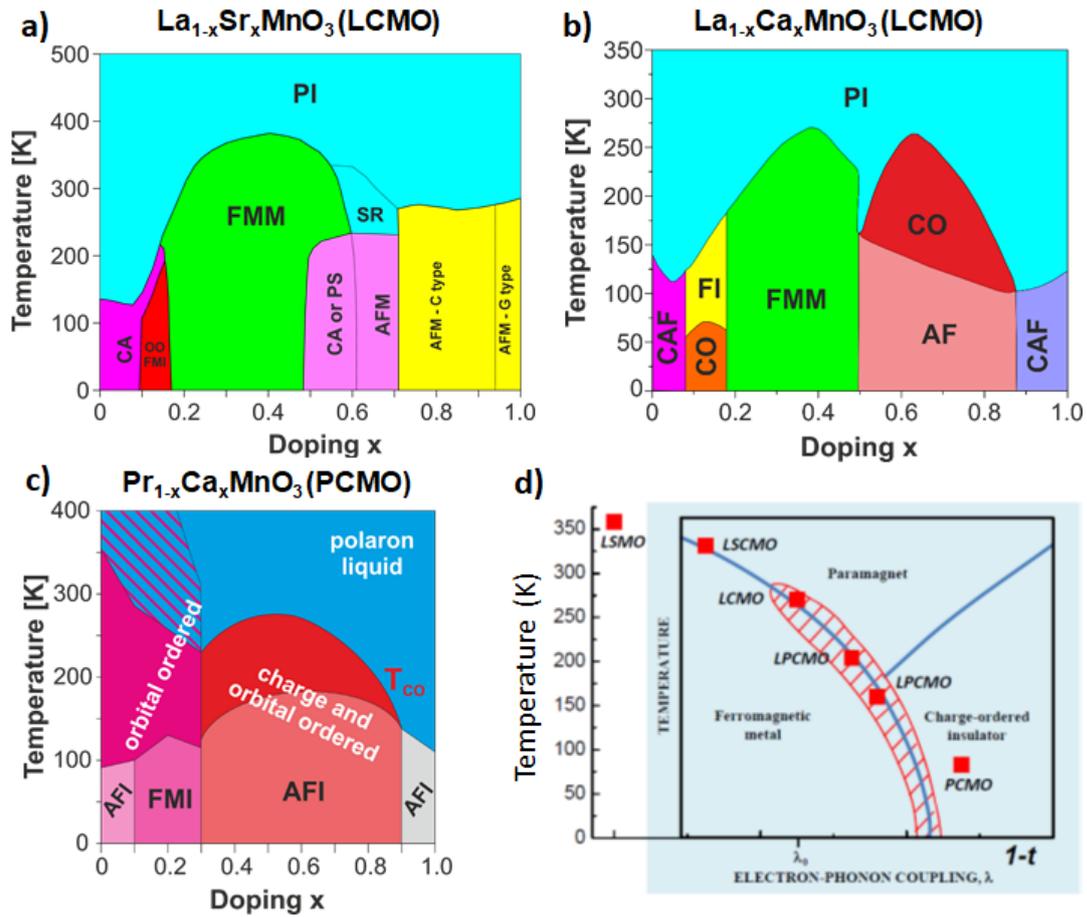

Figure 1: Redrawn phase diagrams of manganites: a) LSMO (according to ref. [20], simplified), b) LCMO (redrawn according to refs. [15a], [15b]; c) PCMO (redrawn according to refs. [21], [22]) and d) a generic phase diagram showing transition temperatures for optimally doped (x=0.33) manganites as a function of electron-phonon coupling (ref. [66]) superimposed by experimental data (red squares) plotted as a function of deviation of tolerance factor (t) from the cubic structure (tolerance factor t=1) (ref. [23]). The hatched area in (d) illustrates electronically inhomogeneous state with coexisting metallic and insulating domains showing CMR.

The ionic picture of doping-dependent coexistence of $Mn^{3+}$ as JT ions with non JT active $Mn^{4+}$ is of course a strong approximation, since due to electron overlap the charge disproportionation is very small. Thus, JT polarons change their properties with doping- or temperature-induced change of couplings, by excitations and due to external fields. For example, an increasing contribution of the double exchange interaction can transform JT polarons to Zener polarons[33],[34],[35]. The correlation of polarons can give rise to ordering:

Charge ordering requires high doping and evolves by "crystallization" to form a periodic polaron crystal. Orbital ordering is due to an ordering of the JT distortion, typically a periodic 90° rotation of the long axis of next neighbor JT displacements. The change of the nature of small polarons from JT to Zener modifies the ordered ground state from a checkerboard CE-type antiferromagnetic (AFM) charge order (JT polaron) to a bond-centered type of charge order with a more complex magnetic order. At a sufficiently high strength of double exchange interaction, the localized or small JT polarons transform into large polarons that can form a ferromagnetic metallic (FMM) ground state[36]. Optical excitation of JT polarons can change both the ordered ground states as well as electron-phonon coupling and interactions of polarons leading to a rich variety of excited state behavior[37],[38]. Changing the density as well as the interaction between polarons can transform a non- or weakly interacting gas of JT polarons to a "polaron liquid" or polaron crystal, i.e. charge order like in PCMO[21].

The importance of the above introduced concept of interacting, tunable QPs within the different ground states as well as their modifications during excitations and phase transitions will be reviewed both by experimental and theoretical studies. The role of QPs for energy conversion at low frequencies $1/\tau=0-10^3$ Hz or long timescales ($\tau=1$ ms–1 hour) is highlighted by experimental studies of local friction and thermal conductivity, showing that they can be tuned by external fields and phase transitions. The high frequency scale $1/\tau=1$ GHz-500 THz or short timescale ($\tau=10$ fs–1 ns) has been addressed by pump-probe reflectivity/resistivity, magnetooptical Kerr effect (MOKE) as well as by photovoltaic measurements. The selected material systems for these studies include thin manganite films with: a) large JT polarons found in $La_{0.7}Sr_{0.3}MnO_3$ (LSMO) and showing ferromagnetic metallic (FMM) ground state; b) a gas of small polarons in $La_{0.7}Ca_{0.3}MnO_3$ (LCMO) and $(La_{0.6}Pr_{0.4})_{0.7}Ca_{0.3}MnO_3$ (LPCMO) resulting in a phase coexistence with competing FMM and CE-AFM phases; c) small polaron liquid and crystal phases found in $Pr_{1-x}Ca_xMnO_3$ x=0-0.5 (PCMO) with charge/orbital ordered insulating ground states with different magnetic order; and d) quasi 2D small polaron crystals found in RP $Pr_{0.5}Ca_{1.5}MnO_4$ (n=1) with charge ordering. The chosen materials are listed along the direction of increasing of strength of electron-phonon coupling (see Fig. 1 d)), accompanied by the increased deviation of the crystal structure from the cubic perovskite[23].

In short, the obtained results evidence an intimate connection between the types of QPs and ordered states, which strongly influence the energy conversion steps of excitation, relaxation and transformation, finally resulting in the demonstration of "hot polaron" photovoltaics. All this emphasizes a huge potential for tuning and controlling energy conversion steps in strongly correlated materials as well as to explore effects of correlated QP's in enhancing the efficiency of energy materials.

## 2. Tailoring the ground state of correlated thin films and heterostructures

Within this section the description of thin film samples as well as of the growth-related modifications of the film properties will be summarized. A high level of growth control is necessary to ensure manganite films and heterostructures with tailored structure and electronic properties. Our focus on thin films is motivated not only because of their relevance for potential energy devices but mainly because of their extreme suitability for tuning the electron-spin-lattice correlations by means of strain and interface engineering. The detailed knowledge of the structure and microstructure of samples as well as of their equilibrium electronic properties is a necessary prerequisite for studies of energy conversion in strongly correlated manganite films.

### 2.1 Film growth and strain engineering.

Epitaxial perovskite oxide thin films have been grown by a variety of different growth methods, including pulsed laser deposition (PLD), ion beam sputtering (IBS) and metalorganic aerosol deposition (MAD)[39]. The advantages of MAD for preparation of various multicomponent oxide films and heterostructures are high oxygen partial pressure $pO_2$~20 % and near-to-equilibrium growth conditions[40], both ensuring preparation of epitaxial multinary oxide films, like $RE_{1-x}A'_xMnO_3$, with desired stoichiometry and high crystalline quality. Moreover, MAD equipped by in situ optical ellipsometry growth control, provides a film thickness control with accuracy better than one unit cell (u.c.) and allows atomic-layer-by-layer epitaxy growth as demonstrated for Ruddlesden-Popper (RP) $Sr_{n+1}Ti_nO_{3n+1}$ (n=4) thin films[39] which cannot be obtained by co-deposition. Remarkably, the layered n=1 and n=2 RP $A_{n+1}B_nO_{3n+1}$ variants of perovskite manganite oxide thin films can be also grown by co-deposition[41], despite their large unit cell. There are special challenges for the choice of substrate, substrate orientation or buffer layers in order to control the growth direction, e.g. uniquely oriented (001) and (010) films. Figure 2.1 shows the growth of a n=1 (010) oriented RP $Pr_{0.5}Ca_{1.5}MnO_4$ thin film on $SrTiO_3$ (110) substrate, where the c-axis is uniquely directed along the [001] direction of the substrate, as demonstrated by x-ray diffraction and transmission electron microscopy[42]. Such films have a significant density of stacking faults which are typical planar defects in RP systems along the c-axis.

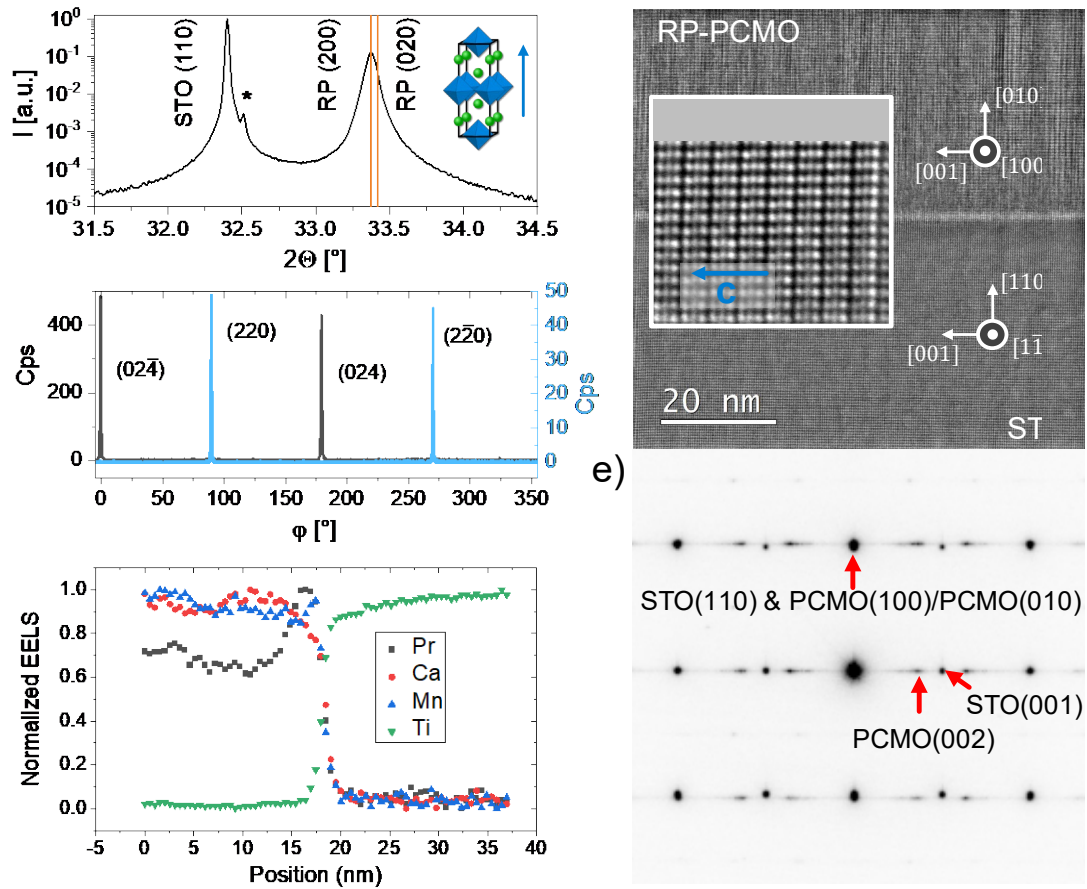

Figure 2.1: Epitaxial thin (010) oriented Ruddlesden-Popper Pr$_{0.5}$Ca$_{1.5}$MnO$_4$ thin film on SrTiO$_3$ (110) substrate. Out of plane (010) and single in-plane c-direction along [001] of the substrates in shown by Θ-2Θ scans (a) and φ-scans (b) using x-ray diffraction. Chemical profile across the interface using electron energy loss spectroscopy (EELS) in TEM showing Pr segregation close to an otherwise sharp interface. (d) Cross-section TEM image of the film in [01$\bar{1}$] zone axis of the substrate. (e) Selective area electron diffraction pattern in the STO [01$\bar{1}$] zone axis shows reflections from the RP PCMO film and the STNO substrate due to the finite size of the SAED aperture. (Ref. [41],[42])

Considering epitaxial growth of manganite films on the typical perovskite substrates, like SrTiO$_3$ (STO) and LaAlO$_3$ (LAO), the lattice-mismatch-induced epitaxy in-plane stress of tensile (+1 %) and compressive (-2 %) origin, respectively, strongly modifies and sometimes degrades electronic properties of thin manganite films[43]. The fundamental reason for strong pressure/strain effects is the extreme sensitivity of magnetic double- and super-exchange interactions[44] on the distortions of perovskite lattice, quantified generally by the Goldschmidt tolerance factor[45] as well as by oxygen octahedral rotation/tilt Mn-O-Mn angle, φ$_{OOR}$[46]. Recently developed strain engineering approaches for oxide films, based on using partially relaxed buffer layers[41],[47],[48],[49] allow to grow and modify oxide films beyond substrate limitations. Thus, the strain state of the film can be tuned independent on the strain actuated by the substrate and it is possible to prepare strain-free thin epitaxial manganite films with intrinsic magnetic and electronic properties as recently demonstrated for the MAD-grown epitaxial $(La,Pr)_{0.7}Ca_{0.3}MnO_3$ (LPCMO) films on LAO-buffered STO(100) substrates[50]. The mechanism for the optimization of magnetotransport in LPCMO/LAO/STO films was found to be the strain-engineering-induced increase of the $\varphi_{OOR}$ angle[50].

**2.2. Emergent polaronic phases in manganite films and heterostructures.**

Another important aspect of the manganite thin film physics is the strong effect of interfaces/surfaces on magnetism and electron transport of thin manganite films, generally summarized within a concept of "dead" layer[51] (DL). DL forms within a few (5-7) monolayers close to the manganite surface/interface due to the symmetry-breaking-induced electronic, orbital and structural reconstructions. As a result, the surface of, e.g., optimally doped $La_{0.7}Sr_{0.3}MnO_3$ (LSMO) film with FMM ground state, acquires unusual AFM insulating properties[52],[53]. The underlying mechanism was found to be a charge transfer from the film bulk to the surface[54], directed to compensate the lack of electrons at the film surface, thus, making it electron-rich and similar to the parent $LaMnO_3$ (LMO) material.

Recent surface studies of different LSMO films by the surface enhanced Raman scattering (SERS)[55],[56], indeed, revealed the JT-distorted LSMO surface with an orthorhombic structure in clear contrast to the rhombohedral volume structure of the optimally doped LSMO film. Moreover, a strong compressive strain ($2 < \varepsilon < 3.4$ %) in thin LSMO/LAO films results in a unique surface behavior with a phase transition at T*=220 K accompanied by an anomalously strong hardening of the JT phonon mode[57]. This surface behavior is very similar to the charge/orbitally ordered (COO) phase transition observed in the volume of the underdoped LSMO films ($0.11 < x < 0.14$)[58], thus, indicating an effectively underdoped surface of compressively strained LSMO. A remarkable similarity between the emerging behavior of surfaces and interfaces can be illustrated by considering the LMO/SMO interfaces with broken inversion symmetry[59],[60]. Namely, electrostatic-mismatch-driven charge transfer (CT) of electrons from the LMO as a "donor" layer to the SMO ("acceptor") within a characteristic CT length scale of $\lambda_{TF}$=2-3 u.c. results in a complex two-phase FM behavior with emergent high-$T_C$ interfacial phase and low-$T_C$ LMO-like volume phase.

We would like to outline a striking relevance of surfaces/interfaces as hosts for JT polarons viewed as short-range-ordered COO phases. Namely, a preferred formation of the insulating COO phase at surfaces and interfaces, is well documented in the literature[61],[62],[63] and can be understood by assuming a weakening of the concurring FM double exchange interaction at defects. Phenomenological Ginzburg–Landau approaches predict formation of a charge-ordered[64] and AFM state[65] within the domain wall, especially when the bulk (domain) phase is located not far away from the FM/AFM boundary in the phase diagram[66]. This is exactly the case of LPCMO films[23] in which such domain walls can be viewed as intrinsic interfaces within the phase separation scenario[67]. Considering the importance of interfaces for energy conversion, especially for the relaxation of QP excitations, exemplified by the role of interfaces (LMO/SMO[59],[60]) in thermal conductivity as well as for transformation of QP excitations into electric current/voltage, e.g., by photovoltaics[42], the knowledge of interface properties and opportunities to control them is of great importance.

### 3. Impact of low energy QPs on friction studied by atomic force microscopy

Friction is a technologically very important phenomenon and has a long history of scientific investigation. However, its microscopic mechanisms are still far from being well understood. Atomic force microscopy, which is sensitive not only to the normal force but also to the lateral one, provides a powerful tool—referred to as friction force microscopy (FFM)—for studying friction locally at the nanoscale at the sliding tip/substrate contact[68]. Most FFM results show a linear dependence of the friction force, $F_f$, on the normal load, $F_n$, in a broad range of $F_n$=5-100 nN, resulting in a nearly constant friction coefficient $\mu=F_f/F_n\sim 0.1$-$0.2$[69].

Over the past 35 years, FFM experiments have been carried out on metals, insulators, semiconductors, and superconductors with the primary aim of uncovering the microscopic mechanisms of surface friction, focusing particularly on the role of electronic and phononic degrees of freedom. Considering strongly correlated electron systems, such as manganites, separating these contributions is often not straightforward because strong electron-phonon coupling leads to formation of new QPs, like JT polarons, which combine both electron and lattice (phonon) degrees of freedom[70]. Even in the metallic state in manganites – with charge carrier concentrations around of $10^{21}$-$10^{22}$ cm$^{-3}$ – polaronic QPs with strongly increased effective mass and low mobility dominate the electron transport, justifying the term "bad metals" for strongly correlated oxides[71]. One may thus expect that such QPs also play a role in dissipation processes, including friction.

### 3.1 Change of friction across phase transitions

Studying how nanoscale friction changes across phase transitions provides insight into how different QP excitations influence it, since such transitions can be induced without altering other surface properties. In particular, it has been shown that the friction coefficient exhibits a step-like decrease by a factor of 2-3 at the metal-superconductor transition in conventional (Nb)[72] and high-$T_C$ superconductors, such as Y-Ba-Cu-O[73] and Bi-Sr-Ca-Cu-O[74]. This has been attributed to the disappearance of the electronic contribution to friction for $T<T_C$, due to the formation of Cooper pairs. In $VO_2$ at the metal/insulator (MI) transition $T_{MI}$=340 K accompanied by a structural phase transition, µ increases by a factor of 2 upon heating across the transition. This change is reversible and reproducible upon both cooling and heating[75],[76]. A similar change in µ has been detected in $VO_2$ by locally probing coexisting insulating (rutile) and metallic (monoclinic) domains at a constant temperature close to $T_{MI}$. Both electronic and phononic (structural) contributions have been proposed to explain the observed friction changes in $VO_2$. A remarkable ability to control friction via electric fields has been demonstrated in p-n junctions fabricated in Si(100) surfaces[77]. A positive bias V=+4 V, which induces hole accumulation in the p-region, results in higher friction compared to the n-region. This excess friction has been attributed to the charging/discharging of the surface caused by the application/release of tip-induced stress.

### 3.2 Change of friction at the metal-insulator transition

The FFM studies on the surface of manganite thin films were particularly aimed to understand the role of polaronic QPs and the interplay between the electron, lattice and spin degrees of freedom in energy dissipation involved in friction. Figure 3.1 shows the studies of friction for an optimally doped (x=0.3) LSMO thin film with $T_C$=338 K and a ferromagnetic metallic (FMM) ground state as well as for x=0.2 with MI transition $T_{MI}$=188 K and $T_C$=220 K[78]. The lateral forces measured in UHV during sliding at a constant scan velocity and under a constant applied normal force, $F_N$=2-25 nN, were obtained from the measured torsion of the cantilever (see Fig. 3.1a)). As one can see in Fig. 3.1b) the friction forces $F_f$ show a linear dependence on the normal load $F_N$=2-25 nN for the temperatures of interest. The temperature dependence of the friction coefficient across the phase transition in the optimally doped LSMO/STO is shown in Fig. 3.1c). One can see the appearance of an excess friction while the sample undergoes the ferro-paramagnetic phase transition at $T_C$=338 K. Simultaneously the metallic state transforms into a kind of polaronic insulating-like state with activated charge transport. Considering the fact that by optimal doping (x=0.3) and within the rhombohedral

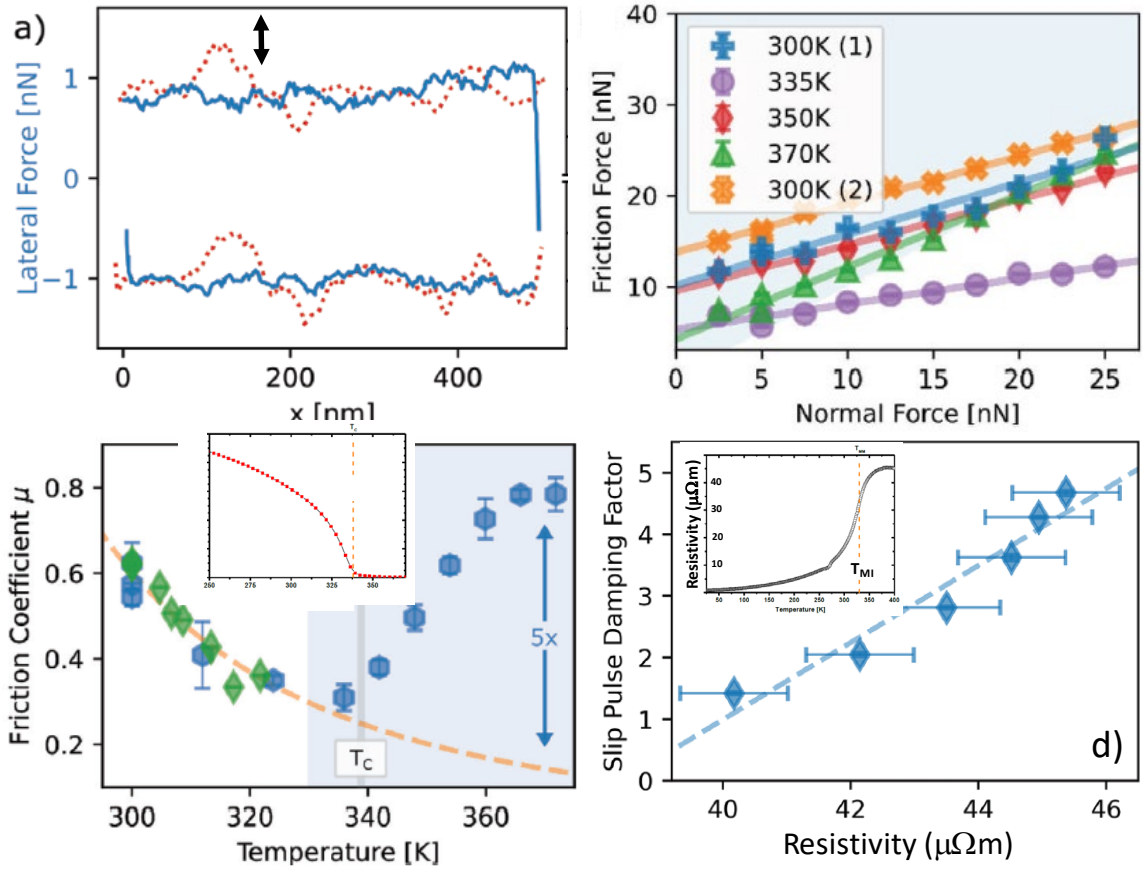

Figure 3.1 Results of friction force measurements on LSMO/STO films: a) Lateral friction force loop (solid lines) and corresponding topography (dotted lines) for the *x* = 0.2 film, measured at an applied normal force of 0.7 nN and scan velocity of 2.5 μm/s, showing a pronounced hysteresis and stick-slip behavior. The arrow indicates the vertical scale of the topography scan. b), c) and d) Data for the *x* = 0.3 LSMO film: b) Friction forces exhibit a linear dependence on the applied normal force across all measured temperatures, both above and below the transition. c) The friction coefficient, determined from the slopes in b) using two different cantilevers (blue hexagons and green diamonds), reveals a marked contrast between the FM metallic state and the PM polaronic conductor state. The dashed line represents an Arrhenius fit to the data. d) The slip pulse damping factor in the polaronic state of the of the *x* = 0.3 film scales with the resistivity near the phase transition. (Redrawn from ref. [78])

LSMO structure static JT distortions are not available[79], the dynamic JT effect plays a dominant role in electron-lattice (polaronic) correlations, resulting in an increase of electrical resistance approaching $T_C$ (see the inset in Fig. 3.1d)).

Friction is generally governed by the stick–slip mechanism[80]. This is also valid for LSMO surface friction as confirmed by the observed thermally activated behavior (Fig. 3.1c)) and the logarithmic dependence of friction coefficient on scan velocity (not shown). Additionally, the absence of changes of adhesion at the phase transition[78] supports the notion that the stick-slip mechanism remains unchanged. The link between excess friction and emergent small polarons in the paramagnetic insulating phase is further illustrated by the observed scaling between the slip pulse damping factor, i.e., the ratio of "blue" data points and "brown" dashed line in Fig.

3.1c), and electric resistivity near $T_C$ (Fig. 3.1d)). These findings support the hypothesis that vibrational slip-pulse excitations couple to phonon degrees of freedom, and via electron-phonon coupling, also to electronic degrees of freedom near the LSMO phase transition.

**3.3 Change of friction via resistive switching by electric fields**

Investigating how friction can be modulated by resistive-switching, induced by applying external electric fields, provides additional insight into the link between dissipation and QPs. Friction force microscopy (FFM), combined with conducting atomic force microscopy (c-AFM), enables such studies. It is known that manganite surfaces can be locally and reversibly switched from a high-resistance-state (HRS) to a low-resistance-state (LRS) by applying voltages V=±4-6 V between a conducting tip and the surface. This process is associated with changes in the JT-induced electronic and structural surface reconstruction[56],[81]. Figure 3.2 shows combined measurements of the friction force by FFM and resistive switching by c-AFM on a MAD-grown $La_{0.55}Ca_{0.45}MnO_3$/MgO(200) (LCMO) film ($T_{MI}$~$T_C$=245 K) in UHV[82]. The resistive state of the manganite switches at a threshold voltage $|V_C| \approx 3$ V[83], while electric current mapping was performed at low voltages (V = 0.1 V<<$V_C$). Figures 3.2 a)- c) present topography and lateral force maps for the initial insulating state (IS), LRS and HRS, respectively. The lateral force variations within each map correlate with the topographic features. A clear reduction in friction by factor of 3-5 is produced by switching while the morphology remains nearly unchanged. This dramatic change in friction force associated with resistive state is most clearly shown by the friction loops shown in Figure 3.2d). Significant hysteresis in the lateral force and thus high friction is observed in the insulating IS and HRS, while the conducting LRS displays almost no hysteresis and much lower friction.

To further explore the relationship between near-surface properties and friction in the LCMO film, the temperature dependence of friction was measured across the MI transition ($T_{MI}$=245 K) in the range of temperatures T=110-300 K. FFM maps were conducted using both Si- and Pt-coated Si cantilevers. Room temperature measurements before and after the temperature series confirmed that neither the specimen nor the c-AFM tip were irreversibly altered. As shown in Fig. 3.2e), friction measured with the Si tip decreases continuously with increasing temperature, consistent with the thermolubricity effect[84]. In contrast, no temperature dependence is observed for the Pt-coated tip; friction remains constant within experimental noise over the entire range of temperatures. Thus, in the case of the LCMO film, contact dynamics entirely controls the temperature dependence of the friction response, while the bulk metal-to-insulator transition remains undetected. This may be explained by the fact that higher energy dissipation occurs in the insulating surface layer (DL), due to localized JT polarons and their strong local electron-phonon coupling. In contrast, the switching-induced LRS, which has more delocalized polarons and is more metallic, exhibits a lower friction. Notably, a thin LSMO film (see Fig. 3.1) shows friction changes across the MI transition, including excess friction attributed to small polarons in the paramagnetic (PM) insulating phase. Such differing behavior between the LSMO and LCMO films likely arises from their very different thicknesses: the much thinner LSMO film (d=5.6 nm) still reflects the MI transition in friction, whereas the thicker LCMO film (d=50 nm) does not. This explanation is supported by FFM studies of LMO/SMO superlattices, all having a terminating LMO layer of varying thickness ($d_{LMO}$=0.7-30 nm)[85]. In these systems, the friction coefficient increases with top layer thickness and then saturates for d > 10 nm, indicating that material properties up to this depth affect energy dissipation at the sliding AFM contact as suggested by the theory[86]. This observation aligns well with the friction change at $T_{MI}$ for the ~5.6 nm LSMO film and its absence in the 50 nm LCMO film. In summary, the change of QP from rather localized small polarons to rather delocalized large band polarons by either temperature- or electric-field-induced phase transitions offers new insights into the friction-related energy dissipation processes. Moreover, this opens up new pathways to tune friction by switchable materials properties.

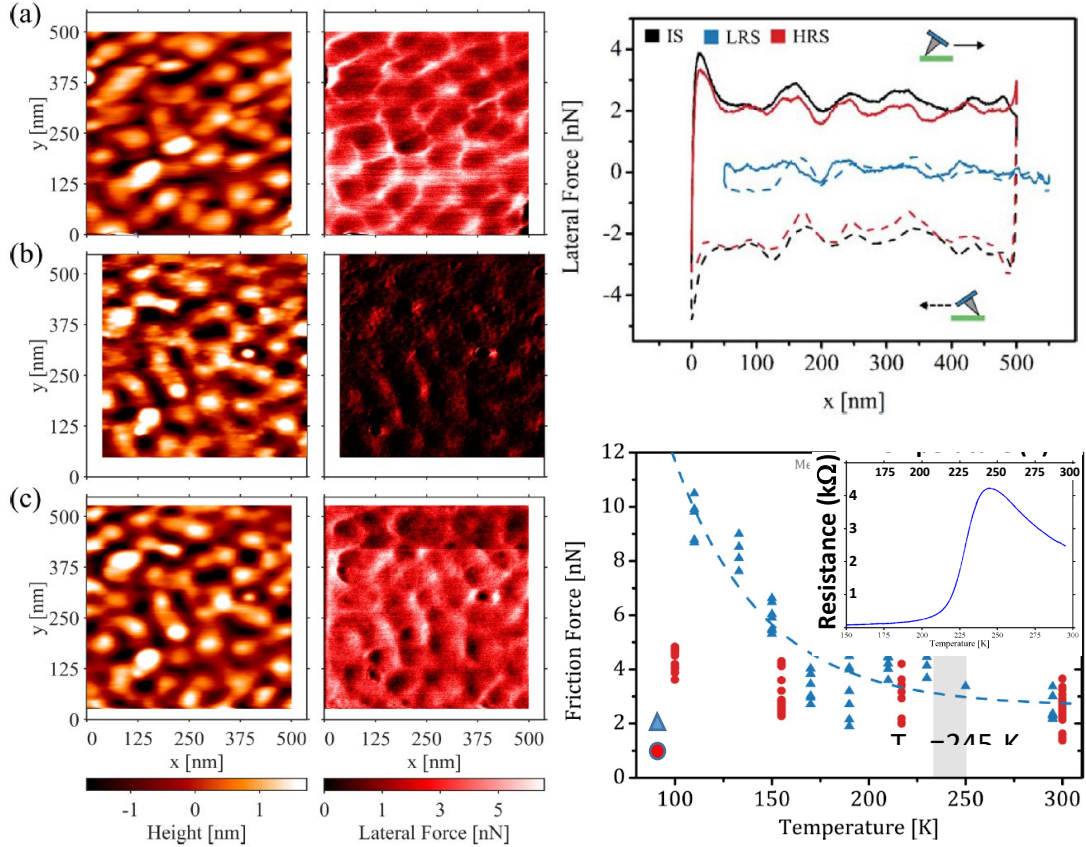

Fig. 3.2 Results of local resistance switching and surface friction experiments on LCMO/MgO film (Ref. [82]). AFM-based measurements using a Pt-coated Si tip: (a–c) Topography (left) and lateral force maps (right) obtained from forward traces measured on (a) the initial insulating state (IS), (b) after resistively switching to the conducting state (LRS), and (c) after resistively switching back to the insulating state (HRS). d) Average friction loops for the insulating states (IS =black; HRS = red) and the conducting state (LRS = blue) obtained with a Pt-coated Si tip. Friction is defined as half the difference between the forward (solid line) and backward (dashed line) lateral force traces. The insulating state exhibits high friction, while the conducting state shows significantly reduced friction. e) Friction force as a function of temperature measured with a silicon (blue triangles) and a Pt-coated silicon tip (red circles). The dashed line represents a thermolubricity model fit to the silicon-tip data with an activation $E_A$=0.2 eV.

### 4. Control of thermal conductivity by QPs and ordering

The transport of heat in materials is controlled by the transport of the available thermal excitations and thus by the formed low energy QPs. In many materials, thermal conductivity can be decomposed into electronic and lattice (phonon) contributions, $\kappa=\kappa_{electron}+\kappa_{phonon}$. This provides fundamental information on the electron and phonon degrees of freedom as well as on the electron-phonon interaction. The interest in thermal conductivity within applied energy research is motivated by the thermal management of electronic devices. For example, the figure of merit of a thermoelectric device, $ZT=S^2\sigma/\kappa$, (temperature T and Seebeck coefficient S) can be optimized by reducing the lattice thermal conductivity $\kappa_{phonon}$. Note, that the electron part of thermal conductivity cannot be reduced without reduction of the related electrical conductivity, σ, since both properties are not independent. In many materials, their connection can be described by the Wiedemann-Franz law $\kappa_{electron}/\sigma=L*T$ ($L=(\pi^2/3)(k_B/3e)^2$ is Lorenz number, e is electron charge). However, in strongly correlated electron systems this law might be violated[87].

### 4.1 Role of phase transition, JT polaronic QPs and octahedral tilts

Thermal conductivity of perovskite manganites is dominated by phonons and was found to be relatively low $\kappa\sim\kappa_{phonon}$=1-5 W/mK[88],[89], reflecting a small value of the phonon mean free path ξ. For example in LSMO, using the kinetic formula, it is estimated to $\xi=3\kappa/C<v>\approx0.4$ nm ~1 unit cell. Here C=130 J/mol*K~$3*10^6$ J/m$^3$K (ref. [90]) is the phonon specific heat per volume and <v>=5900 m/s (ref. [91]) is the average phonon velocity or Debye (sound) velocity. This indicates diffusive phonon transport in manganites, substantiating its high sensitivity to lattice distortions originated by octahedral tilts and JT polarons[88]. Since electron-phonon coupling and polaronic QPs type as well as their concentration can strongly vary at temperature- and field-induced phase transitions, this can have a large impact on thermal conductivity[92].

Figure 4.1A) shows the results of magnetic (top panel), electric (middle), and thermal

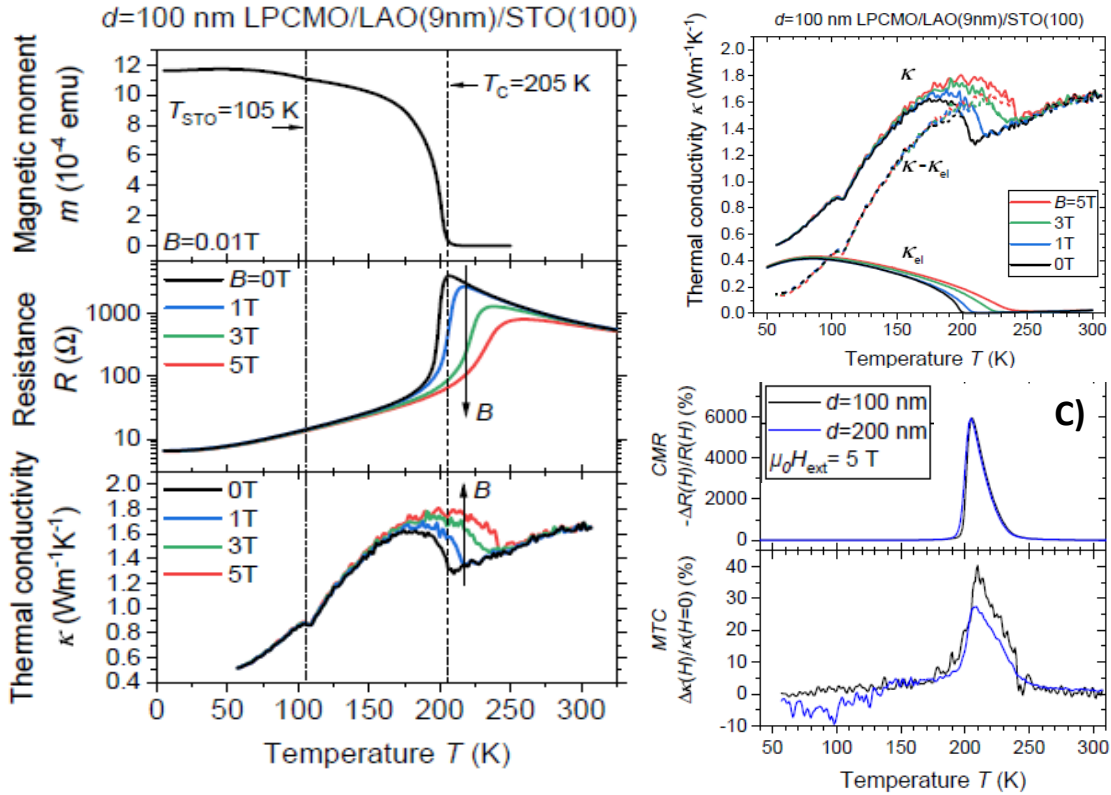

Figure 4.1: Thermal and electric transport in an LPCMO/LAO/STO film: A) Temperature dependences of magnetic moment (top), resistance (middle) and thermal conductivity (bottom); B) electronic and phononic parts of thermal conductivity and C) Comparison between CMR and magnetothermal conductivity (MTC).

transport (bottom) measurements carried out by using the 3ω technique[93] on an epitaxial 100 nm thick LPCMO film grown on LAO-buffered STO with relaxed epitaxy strain[50]. One can see coupled para-ferromagnetic and MI transitions at $T_C=T_{MI}=205$ K. Thermal conductivity decreases with decreasing temperature in the PM insulating phase and abruptly increases as the film enters the FMM state. A similar increase of κ is observed in applied magnetic fields close to the phase transition (see Fig. 4.1 A), whereas far away from $T_C$ magnetic field has no effect on κ as well as on the electrical resistance. In Fig. 4.1B) one can see temperature dependencies of the electronic part of thermal conductivity $\kappa_{el}=L*T/\sigma$ calculated from the measured electrical conductivity $\sigma(T)=1/\rho(T)$ using the Wiedemann-Franz law. The thermal conductivity due to phonons, $\kappa_{ph}=\kappa-\kappa_{el}$, also shown in Fig. 4.1 B), dominates thermal conductivity for high temperatures, T=120-300 K, and shows a significant change at the phase transition driven by temperature and/or magnetic field. At low temperatures deeply in the FMM state (T<100 K) the electronic contribution prevails the lattice one because of freezing of phonons.

An apparent correspondence between the CMR effect and the so called magnetothermal conductivity[94] MTC=[κ(T, B=5 T)-κ(T, B=0)]/κ(T, B=0) can be seen in Fig. 4.1 C). Namely, an applied magnetic field B=5 T for temperatures in the vicinity of phase transition results in a 6000 % suppression of electrical resistivity and yields up to 40 % increase of thermal conductivity dominated by the lattice contribution. CMR as a magnetic-field-induced insulator-metal transition in LPCMO[13] is intimately related to a metamagnetic transition[67] accompanied by annihilation of short-range ordered static JT distortions (small polarons) and transformation into delocalized band polarons. Both temperature- and field-driven increase of $\kappa_{ph}$ in LPCMO in Fig. 4.1 illustrates nicely a pronounced enhancement of the lattice order at the phase transition which can be viewed as a disorder/order transition coupled to polaronic degrees of freedom, additionally supported by the suppression of JT phonon bands in Raman spectra[95]. Note that the lattice symmetry of LPCMO does not change at $T_C$ and the structure remains orthorhombic[67].

In Fig. 4.2, a remarkable generic similarity between the temperature dependences of thermal conductivity in an LPCMO film and of the Mn-O-Mn bond angle, $\varphi_{OOR}$, measured by neutron diffraction in bulk $La_{0.75}Ca_{0.25}MnO_3$ (LCMO)[46] is shown. Despite differences in $T_C$ (240 K for LCMO and 205 K for LPCMO) as well as in the sharpness of the phase transition, the bond angle, $\varphi_{OOR}$, and lattice part of thermal conductivity, $\kappa_{phonon}$, both reveal a sudden increase in the metallic phase. This illustrates a tight connection of thermal conductivity and octahedral tilt angle. Thus the tilt angle can be viewed as a generic structural control parameter in perovskite manganites because it controls the overlap integral between Mn3d and O2p states and thus the electronic bandwidth D. The collapse of rather delocalized large polarons into localized small polarons and their JT lattice disorder takes place at a threshold of the coupling parameter $\lambda_c > 1.5$, where $\lambda = E_p/D$ and $E_p$ denotes the polaron formation energy[96],[97]

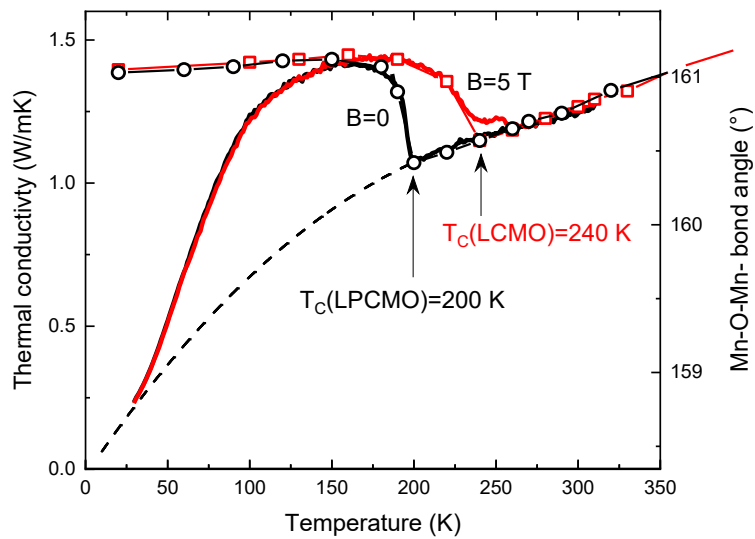

Figure 4.2 Left scale: thermal conductivity of an LPCMO film in ambient (B=0, black) and applied field B=5 T (red). Right scale: octahedral Mn-O-Mn tilt/rotation angle, $\varphi_{OOR}$, as a function of temperature for LCMO bulk (red squares, ref. [46]) and the reconstructed $\varphi_{OOR}(T)$ dependence for LPCMO film (black circles) using the fact that this film in applied field of 5 T shows $T_{MI}$=240 K as bulk LCMO. The dashed line is an extrapolation of the high temperature lattice contribution to thermal conductivity using a $T^3$ law.

### 4.2 Impact of charge ordering

PCMO as a narrow-bandwidth perovskite manganite with orthorhombic structure and small Mn-O-Mn bond angle $\varphi_{OOR}$=154° (ref. [46]) possesses a ground state of an AFM CO insulator with $T_{COO}$=200-220 K (ref. [22]). $PCMO_t/STO_t$ SLs with thickness t=1-25 nm have been grown by PLD and MAD techniques on STO(100) substrates with the main aim to study the impact of CO transition on the phonon thermal conductivity in very thin PCMO layers[98]. Thermal conductivity has been measured by 3ω technique in the temperature range of T=40-310 K using an isolating STO top layer. Interface engineering of SLs has been applied within MAD by using SrO layers grown on PCMO/STO interfaces in order to elastically decouple PCMO and STO layers[99] and relax the epitaxial strain. As one can see in Figure 4.3 a) this interfacial octahedral decoupling resulted in an almost perfect growth of the whole SL consisting of 32 bilayers at the cost of formation of stacking faults at the interfaces by means of SrO monolayers (see Fig. 4.3 b)). Without SrO interface engineering SLs with very thin layers t=1.7-3.5 nm possess macroscopic wave-like defects due to the strain-induced bending of layers[98].

Figure 4.3 c) and d) show the temperature dependences of thermal conductivity in SLs. The strong peak at a temperature $T_P$=220-240 K observed in SLs with t>3 nm agrees well with the evolution of the CO phase transition in PCMO at $T_{CO}$=220 K. An indication for a hysteresis in κ(T) curves during cooling and warming can be also seen in Fig. 4.3 d) in agreement with the 1st order nature of CO phase transition. SLs with thinnest layers t=1.75-1.85 nm do not show such peaks and display smaller values κ(T) similar to a behaviour found in amorphous materials. Evidence for the existence of CO in the PCMO films is shown by electron diffraction pattern in Fig. 4.3 e) and d) with superstructure peaks (marked by arrows) [0, (n + 1)/2, 0], for

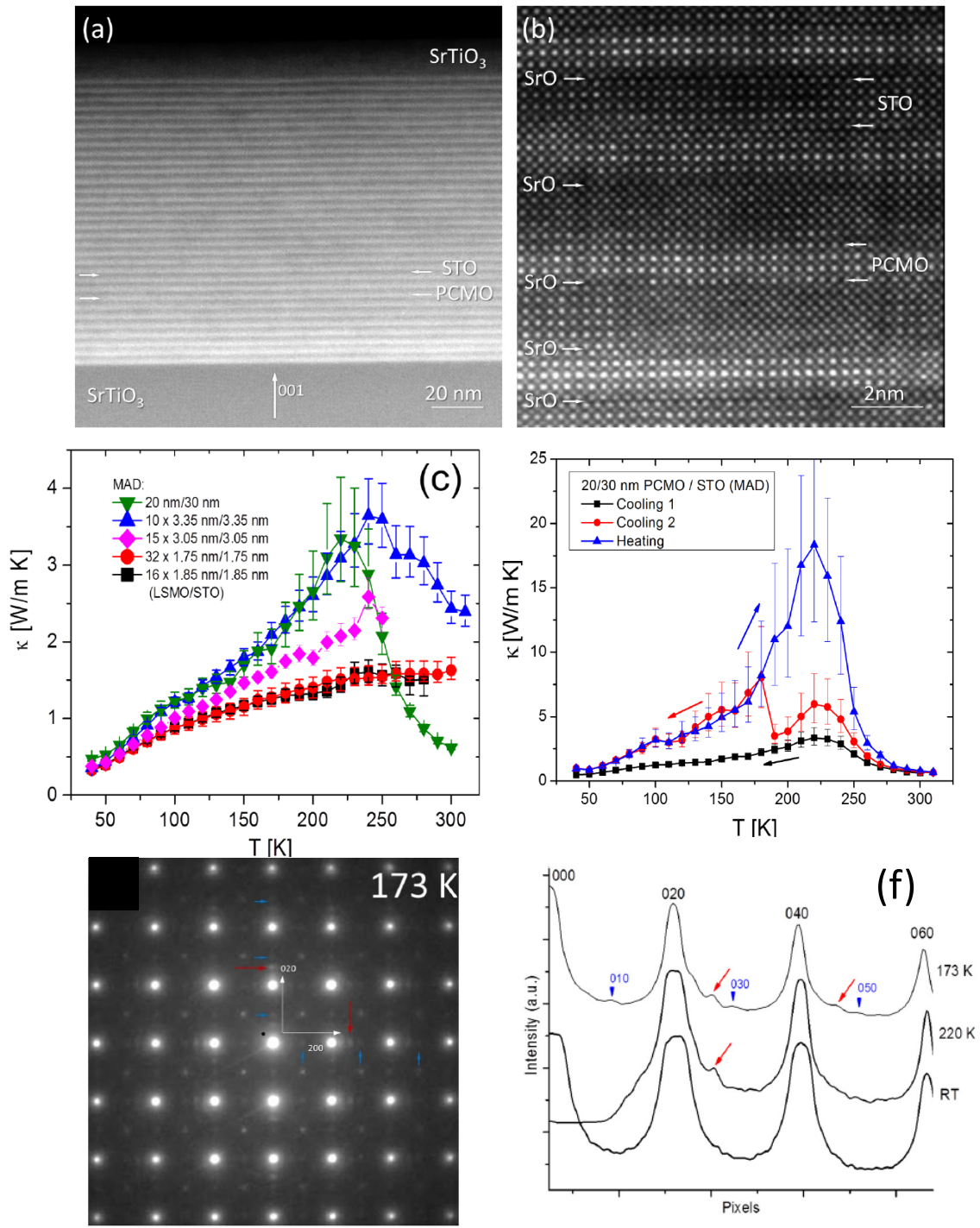

Figure 4.3: a), b) HAADF-STEM images of interface-engineered PCMO/STO superlattices; c) and d) temperature dependences of thermal conductivity for different SLs showing a peak in the vicinity of $T_C$=220-240 K; e) Selected area electron diffraction (SAED) measured in the PCMO/STO bilayer; f) the SAED scans along the [0k0] zone axis at different temperatures showing formation of CO superstructure peaks below T~220 K. Redrawn from ref. [98].

n = 0, 1, 2 ,. at temperatures $T<T_{CO}\sim 220$ K, in close agreement with the peak in $\kappa(T)$ in Fig. 4.3. The maximum of cross-plane thermal conductivity in strongly correlated PCMO$_t$/STO$_t$ SLs close to the CO phase transition in PCMO is interpreted as a phonon softening at the phase transition which is absent in ultrathin films due to suppression of the CO/OO phase transition.

### 4.3 Effects of magnetic order/disorder

By stacking dissimilar manganite layers in an SL one gets a promising opportunity to study the effects of reduced layer thickness, strain and electronic/structural transformation at the interfaces on the emerging properties, like magnetism[59]. B-site ordered double perovskite $La_2MnCoO_6$(LMCO) is a hard FM material due to the underlying superexchange interaction[100], saturation magnetization 6 $\mu_B$/f.u. (f.u.: formula unit) and perpendicular magneto-crystalline anisotropy (PMA)[101]. In contrast, LSMO as a classic double-exchange material is a soft FM with bulk $T_C$=360 K and in-plane easy magnetic axis. Recently, we have observed a peculiar exchange-spring magnetic behavior in LMCO/LSMO SLs[102]. As shown in Figure 4.4 exchange-spring magnetic behavior results in a spin reorientation transition of soft LSMO layers at the Curie temperature of LMCO which was increased up to $T_C$(LMCO)=260-280 K in SLs compared to bulk samples having $T_C$=225 K[100]. This spin reorientation transition yields a crossover from the in-plane (IP) magnetic anisotropy originating from the LSMO to the PMA governed by LMCO. By applying in-plane magnetic fields one can change back to the IP anisotropy and, thus, control magnetic disorder in SLs by forming Neel domain boundaries at (close to) the interfacial regions.

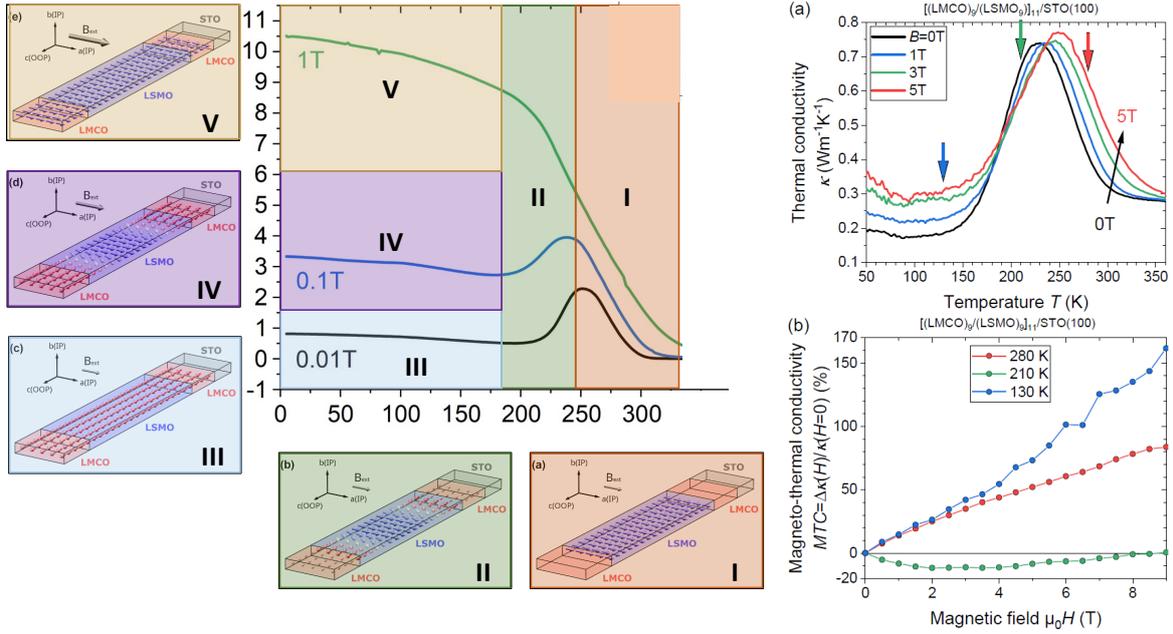

Figure 4.4: Left panel: temperature dependence of magnetic moment in a $LMCO_9/LSMO_9$ superlattice reflecting the exchange-spring magnetic behavior with spin reorientation of soft LSMO layers at the Curie temperature of hard LMCO layers ($T_C$=260 K); spin configurations are shown by sketches with colors corresponding to the temperature regions (I-V). Right panel: temperature dependence of thermal conductivity of the $LMCO_9/LSMO_9$ SL measured in applied magnetic field B=0-5 T (top) and magnetic field dependence of magnetothermal conductivity, MTC=[κ(T, B=5 T)-κ(T, B=0)]/κ(T, B=0) measured for three different temperatures (bottom). Redrawn from ref. [102]

The cross-plane thermal conductivity in LMCO$_9$/LSMO$_9$ SLs, measured by 3ω technique for temperatures T=40-350 K and magnetic fields B=0-9 T, shown in Fig. 4.4 b) reveals a complex behavior with a maximum in the vicinity of spin reorientation transition and shows a strong influence of magnetic field on the phonon transport. The conductivity κ(T) increases as LSMO becomes FMM for T<T$_C$(LSMO)=300 K and then starts to decrease at T$_C$(LMCO)=260 K in good agreement with spin reorientation and enhancement of magnetic disorder within SL. Far away from the phase transitions both in the LMCO and LSMO thermal conductivity in SL reveals very low values κ~0.2 W/mK, which are significantly smaller than those observed for LSMO κ≈2 W/mK, but are typical for other double perovskites[103]. Considering the cross-plane measurement geometry and serial connection of LSMO and LMCO thermal resistances, thermal conductivity of these SLs is indeed determined solely by LMCO layers having a much larger thermal resistance than LSMO. Applied magnetic fields for high (T≥T$_C$(LSMO)) and low (T<<T$_C$(LMCO)) temperatures result in a significant increase of thermal conductivity by factor of 2-3 due to the increase of magnetic order in the LSMO and LMCO layers, respectively. Interestingly, for intermediate temperatures 180 K<T<250 K where the spin-reorientation-induced magnetic disorder is maximal, the applied field leads to a sizable decrease of thermal conductivity, yielding a negative MTC (see Fig. 4.4 c)).

These results illustrate the versatility of SLs and a possibility for magnetic tunability of their thermal transport. Moreover, the SLs containing double perovskite layers show extremely low values κ~0.2 W/mK for crystalline materials originated, probably, due to a periodic modulation of Mn-O-Mn bond angle φ$_{OOR}$ along the growth direction[102].

### 4.4 Effects of bond angle in LMO/SMO superlattices

A direct modification of octahedral tilt and thus of Mn-O-Mn bond angle by SL design and its influence on thermal transport[104] as well as on emergent magnetism[59],[60] has been demonstrated in LMO/SMO SLs grown by MAD. Figure 4.5 shows correlations between the SL period, interfacial structure and cross-plane thermal conductivity measured by all-optical thermal transient reflectivity (TTR)[105] for T=100-350 K in ambient magnetic field. The SLs consist of digital units of [(LaMnO$_3$)$_m$/(SrMnO$_3$)$_n$]$_K$ (LMO/SMO) with bilayer number K=10-20. The thickness of individual layers is m,n=3-20 u.c. and ratios m/n=1 and 2 have been realized (see details in [59],[60],[104]). The main result is the correlation between the local Mn-O-Mn bond angle $\varphi_{OOR}$ and thermal conductivity. The iDPC-STEM (see Fig. 4.5 c and d)) is able to visualize the oxygen atoms and to estimate the Mn-O-Mn bond angles[106],[107] and reveals a characteristic difference between the SLs with m/n=2 and those with m/n=1. Namely, the LMO layers within m/n=2 SLs possess $\varphi_{OOR}$=165°±3° typical for the rhombohedral LMO[108], whereas the m/n=1 SLs show unusually large for LMO $\varphi_{OOR}$=180°±3° typical for cubic perovskites. The SMO layers in both types of SLs display large values $\varphi_{OOR}$=180°±3° characteristic for bulk SMO with cubic structure[109]. Thus, we conclude that a large MnO$_6$ octahedral interfacial tilt mismatch is present in all SLs with m/n=2, whereas the interfaces in the SLs with m/n=1 are tilt-mismatch-free.

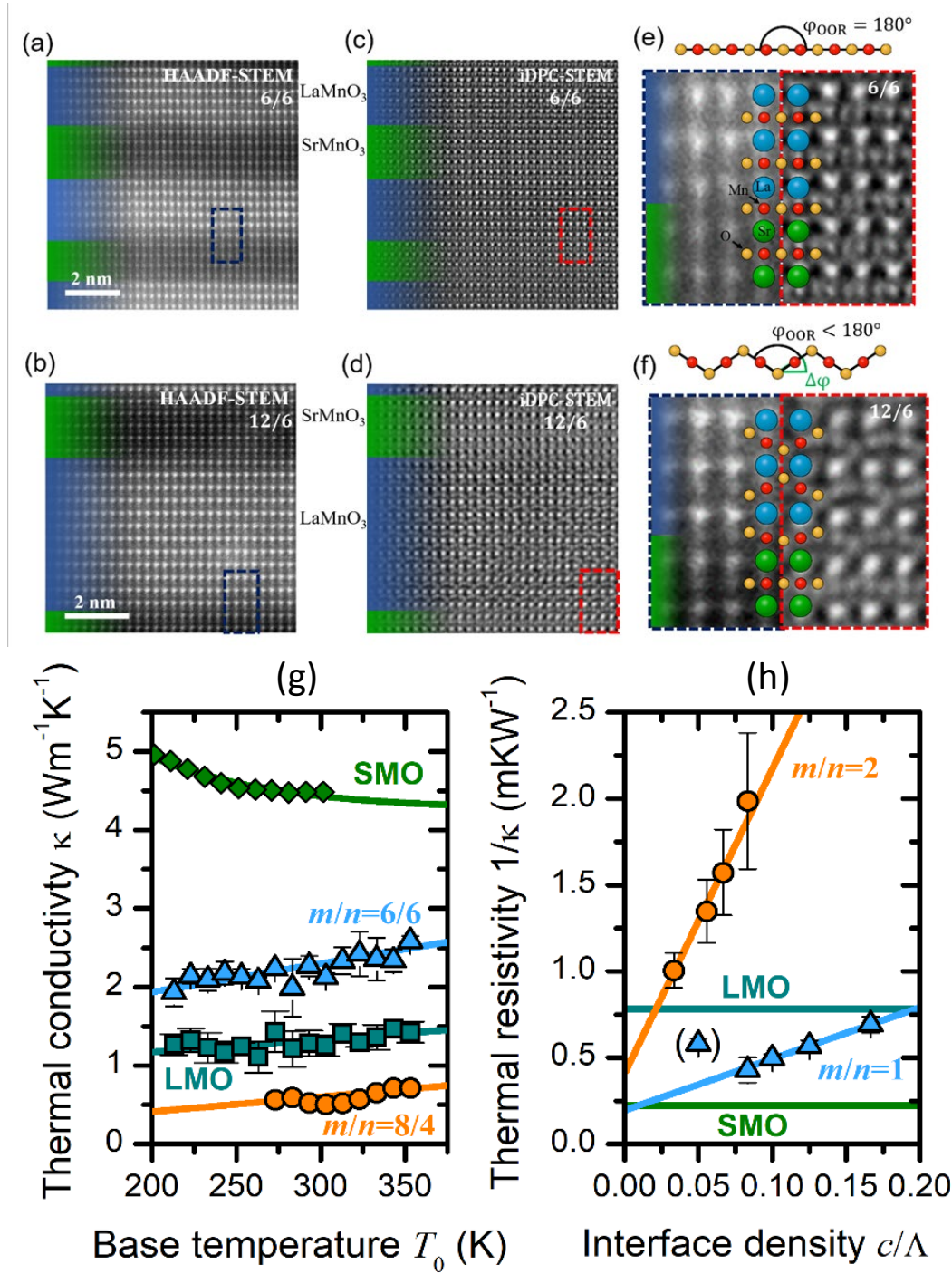

Figure 4.5 High resolution transmission electron microscopy ((a)-(f)) and phonon thermal transport ((g), (h)) data for LMO/SMO superlattices as well as for the LMO and SMO single films with thickness of 40 nm. Redrawn from ref. [104].

Thermal resistivity (see Fig. 4.5 h)) depends linearly on the interface density, thus fitting well the multilayer model[110], i.e., $1/\kappa=1/\kappa_V+2/h\Lambda$ with $1/\kappa_V$ is the volume thermal resistivity and $h$ is interfacial conductance. Two features are apparent: 1) thermal resistivity is dominated by the interface contribution; the latter seems to be strongly different for the SLs with different m/n ratio. Indeed, thermal interface (Kapitsa) conductance calculated from the slope of linear fits in Fig. 4.5 h) was found to be very different in these two types of SLs: *h(m/n=1)=1.8 GWm$^{-2}$K$^{-1}$>>h(m/n=2)=0.3 GWm$^{-2}$K$^{-1}$*. Thus, interfacial conductance is by factor of 6 larger for m/n=1 SLs than that for m/n=2 SLs; and 2) the volume thermal resistivity, $1/\kappa_V$, estimated by extrapolation of fit lines to $1/\Lambda=0$, is also very different in these two types of SLs. Namely, for the m/n=1 SLs $1/\kappa_V$ is very close to the thermal resistivity of SMO, meaning that LMO layers in the SL behave like SMO layers or in other words thermal conductivity of LMO is strongly increased up to the level of SMO. In contrast, the m/n=2 SLs display $1/\kappa_V$=0.43 mK/W, which fits quite well to the weighted average of the LMO and SMO contributions: *$\kappa_V=(m+n)/(m/\kappa_{LMO}+n/\kappa_{SMO})$*. Clearly, the presence of octahedral tilt/rotation mismatch at the interfaces within the m/n=2 SLs, characterized by significantly different Mn-O-Mn bond angles in LMO ($\varphi_{OOR}$=165°) and SMO ($\varphi_{OOR}$=180°), leads to the reduction of interfacial (Kapitsa) conductance as well as to the overall reduced thermal conductivity. The absence of octahedral mismatch at LMO/SMO interfaces in the m/n=1 SLs, where both LMO and SMO possess optimally large $\varphi_{OOR}$=180°, leads to the increase of interfacial conductance as well as of the overall thermal conductivity. Note, the surprisingly large *h=1.8 GWm$^{-2}$K$^{-1}$* interfacial conductance in the m/n=1 SLs, which is comparable with the record value of 4 GWm$^{-2}$K$^{-1}$ observed in the metallic Al-Cu interfaces[111]. All of this points to the octahedral tilt/rotation angle $\varphi_{OOR}$ as a generic structural control parameter for magnetism, electron and thermal transport in perovskite manganites and superlattices, since it controls the overlap integral *t* between O2p and Mn 3de$_g$ states. Compared to *t* = 0.65 eV of the ideal cubic structure ($\varphi_{OOR}$=180°) it can drop down by up to 30%[112], thus undermining ferromagnetic double exchange and inducing a collapse of large band polarons into self-localized small polarons (see e.g. ref. [97]). In particular, LMO layers in m/n=1 SLs with thickness 3-10 u.c. and large $\varphi_{OOR}$~180° possess an FM insulating ground state with unusually large $T_C$=300-310 K[60] that seems to be incompatible with the presence of small JT polarons. In contrast, the LMO layers in m/n=2 SLs with relatively small $\varphi_{OOR}$~165° have reduced values of $T_C$=160-270 K due to suppression of double exchange interaction[44] showing thermal behaviour qualitatively similar to that observed for small JT polaron QPs.

## 5. Band structure and optical excitation of small polaron QPs

In strongly correlated electron systems, the origin of energy bands of electrons can be very different compared to conventional semiconductors. Electrons in weakly correlated materials can be described as weakly interacting QPs moving in a rigid band structure, while the atomic and spin structure are well described by their mean values. Adding or removing electrons does not affect the atomic or spin structure in a fundamental way. In contrast, adding electrons to a strongly correlated material results in new composite QPs such as polarons, excitons. Changing their concentration will modify their interactions and thus the electronic ground state. Or the additional electron may be attached to phase boundaries present in the material. A description of the material must consider the cooperative effects on the band structure of the electrons. For example, as introduced in section 6, optical excitations can lead to phase transitions, accompanied by the change of electronic structure as well as by changes of the nature of QPs[113],[114],[115],[116]. As shown in the following sections, this will have a strong impact on photovoltaic or other types of energy conversion, thus, opening up new pathways for e.g. stabilization of hot excited states and energy harvesting.

Our focus lies on a specific material class chosen as a representative example, namely 3d-transition-metal-oxide perovskites, namely manganites with polarons that produce ground states with various orbital, charge and spin structures. Many of the driving mechanisms can already be rationalized considering only a pair of Mn sites and their oxygen octahedra. Due to the electron-phonon coupling, an electron in the degenerate set of $e_g$ electrons of a transition metal distorts the surrounding oxygen octahedron, which in turn splits the $e_g$ doublet into a filled lower and an empty upper orbital. Neighboring polarons tend to arrange themselves either with opposite orbital polarization and equal spin polarization or with opposite spin polarization but equal orbital polarization[117]. Within a pair of a polaron $Mn^{3+}$ and an empty site $Mn^{4+}$, the electron tends to delocalize over both sites and form a Zener polaron with aligned spin and orbital polarization. The subtle balance of these effects is at the root of the complex phase diagrams[15a],[20],[21],[22],[23],[66] and the resulting effects such as colossal magnetoresistance in manganites[13].

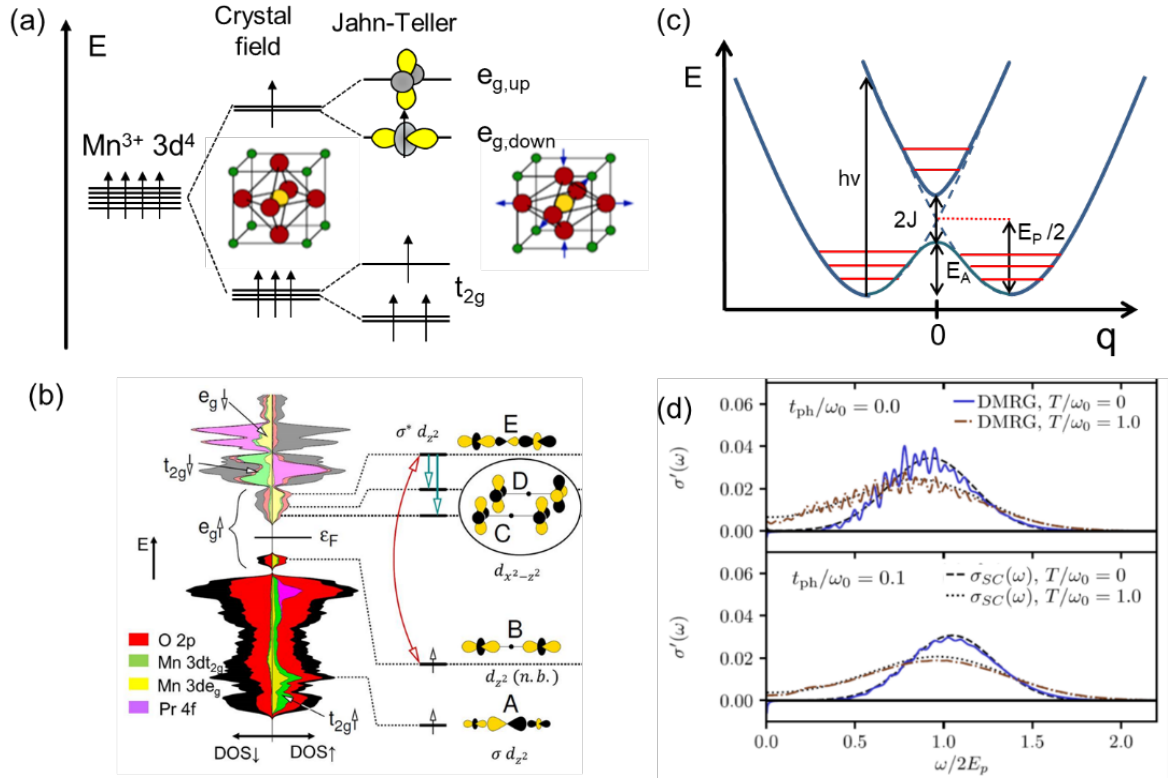

Figure 5.1: Electronic states, band structure and quasiparticle excitations in materials with strong electron-phonon correlations. (a) Scheme of lift of the degeneracy of 3d states in Jahn-Teller (JT) active transition metals by crystal field and JT effect for the example of $Mn^{3+}$. This type of coupling of the electron state to lattice distortion is the origin of polaronic effects in many 3d transition metal oxides. b) Density of states of half doped $Pr_{0.5}Ca_{0.5}MnO_3$ obtained from density-functional calculations with corresponding Mn-O-Mn molecular orbitals of a JT dimer[118],[121]. Majority and minority spin directions are shown for total density of states (black), projections onto O p (red), as well as Mn $t_{2g}$ (green) and Mn $e_g$ (yellow) orbitals of the spin-up manganese atoms as stacked plot. The JT splitting and covalent Mn-O-Mn interaction of the Mn $e_g$ states governs the band structure close to the Fermi energy, leading to occupied σ-nonbonding (B) and unoccupied σ-antibonding (E) Mn-O-Mn states as well as unoccupied orbitals (C, D) with δ character. (c) Energy potential surface as a function of the atomic displacement coordinate q in the two-site Holstein model for small polarons. (d) DMRG calculation of the optical conductivity of a small polaron in the Holstein model at different temperatures for two different phonon bandwidths $t_{ph}/\omega_0$ and strong coupling $\lambda$=4.5 (Ref. [119]).

A low-energy optical excitation of a JT polaron can be described in different frameworks as summarized in Figure 5.1. It lifts an electron from the occupied lower JT orbital shown in Fig. 5.1 a) into the empty orbital of the neighboring site. Such a process can be described in the simple model of a Jahn-Teller dimer, where the resulting molecular orbital scheme together with the DFT calculation of half doped system $Pr_{0.5}Ca_{0.5}MnO_3$ is shown in Figure 5.1b. The excited electron-hole pair opposes the existing orbital polarization on either site, which in turn modifies and often reverts the octahedral distortions. Thermal fluctuations increase the intensity in the low-frequency side of the JT excitation, which may be described either by a low-energy shift or an additional low-energy feature in the absorption spectrum. The effect is due to the increase of the optical transition matrix elements and a simultaneous decrease of the absorption energy with decreasing JT distortion. The thermal shift towards lower absorption frequency is a generic effect that can already be described in the Holstein dimer model for small polarons. Figure 5.1 c) shows the energy potential surface of the Holstein two site model[120], with an optical excitation from the ground state. The typical skewed Gaussian shape of the calculated optical conductivity at higher temperature is shown in Fig. 5.1 d).

Figure 5.2 shows the electronic band structure of the half-doped $Pr_{0.5}Ca_{0.5}MnO_3$ system and the Ruddlesden-Popper variants n=1 and n=2. For all of them the density of states around the Fermi level is dominated by the JT type of electron-lattice correlations, forming a lower and an upper 3d-$e_g$ bands, where the band splitting is of polaronic nature. On-site Coulomb interaction and magnetic exchange additionally shifts the energy levels of the 3d states with respect to the O2p states. They all show a small polaron absorption feature in the optical conductivity due to dipole allowed transitions from an occupied non-bonding $\sigma$ to an antibonding non-occupied states of $\sigma$ and $\delta$ character (compare Figure 5.1b). Here, an optical excitation of polarons (will be discussed in more detail in sec. 6) induces an occupation of the upper $e_g$ state, which does not match the JT distortion of the involved octahedral[121]. After optical excitation of polarons a coupled reorganization processes of electron, spin and lattice degrees of freedom set in and lead to a variety of different relaxation and thermalization

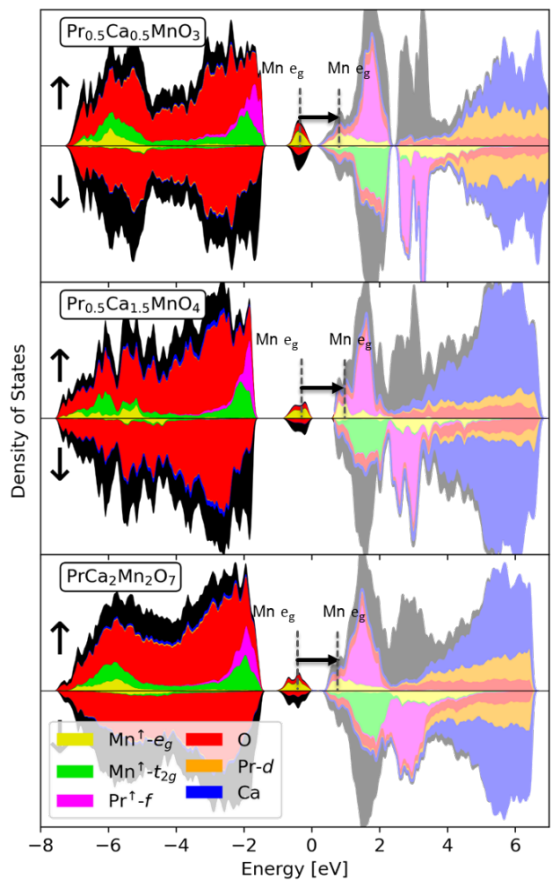

Figure 5.2: Density of states of half doped $Pr_{0.5}Ca_{0.5}MnO_3$, the n=1 RP system $Pr_{0.5}Ca_{1.5}MnO_4$ and the n=2 RP system $PrCa_2Mn_2O_7$ obtained from density-functional calculations. Majority and minority spin directions are shown as well as total density of states (black), projections onto O p (red), as well as Mn $t_{2g}$ (green) and Mn $e_g$ (yellow) orbitals of the spin-up manganese atoms as stacked plot. All three systems have in common the JT splitting of the Mn $e_g$ states governing the states close to the Fermi energy with minor shifts in energy. The dipole-allowed optical transition of the small polaron from the lower to the upper $e_g$ state is indicated.

dynamics which can be strongly affected by the type of ordering[22].

Figure 5.3 shows the small polaron optical absorption in different manganites, including the charge ordered $Pr_{1-x}Ca_xMnO_3$ x=0.35 (a) and x=0.5 (b)[122], the orbital ordered $Pr_{1-x}Ca_xMnO_3$ x= 0.1 (c)[22], the half-doped RP of $Pr_{0.5}Ca_{1.5}MnO_4$ (n=1) (d)[42] and $PrCa_2Mn_2O_7$ (n=2) (e) as well as the $((La_{1-y}Pr_y)_{1-x}Ca_xMnO_3)$ (y=0.4, x=0.3) (f) that shows an insulator to metal transition[123]. Figure 5.3 (d) shows the fit of the analytical solution of the two-site Holstein model as derived by Reik and Heese[124]. The maximum of the skewed Gaussian is connected to the polaron formation energy $E_p$ by $\hbar\omega_{max} \approx 2E_p$, with a correction that leads to an increasing redshift with the thermal broadening[122]. The additional shoulder at the low energy side of the peak can be fitted by an additional skewed Gaussian and is due to a maximum of spectral weight for transitions from the top of the lower to the bottom of the higher energy potential surface due to thermally activated hopping (see Figure 5.1 d)). The skewed Gaussian dependence on photon energy is typical for JT type of small polaron absorption. It is not only observed in many manganites[125],[126], but also in other strongly correlated perovskite oxides such as nickelates and cuprates[127]. Only the low-doped orbital ordered system in Figure 5.3 c) exhibits an optical bandgap of about 1.1 eV. Its origin was first attributed to a Mott-Hubbard splitting due to a strong Coulomb repulsion of the rather localized electrons with high density in the $e_g$ bands at low doping, but later found to be dominated by the $e_g$-$e_g$ splitting due to a cooperative JT type of orbital ordering[128] (see also Sotoudeh et al[118]). Similar absorption features resembling small polaron effects were observed in ruthenates and discussed in terms of Mott-Hubbard physics[129]. However, a detailed structural analysis recently confirmed that JT-related orbital order is present in ruthenates and could be even involved in driving superconductivity[130].

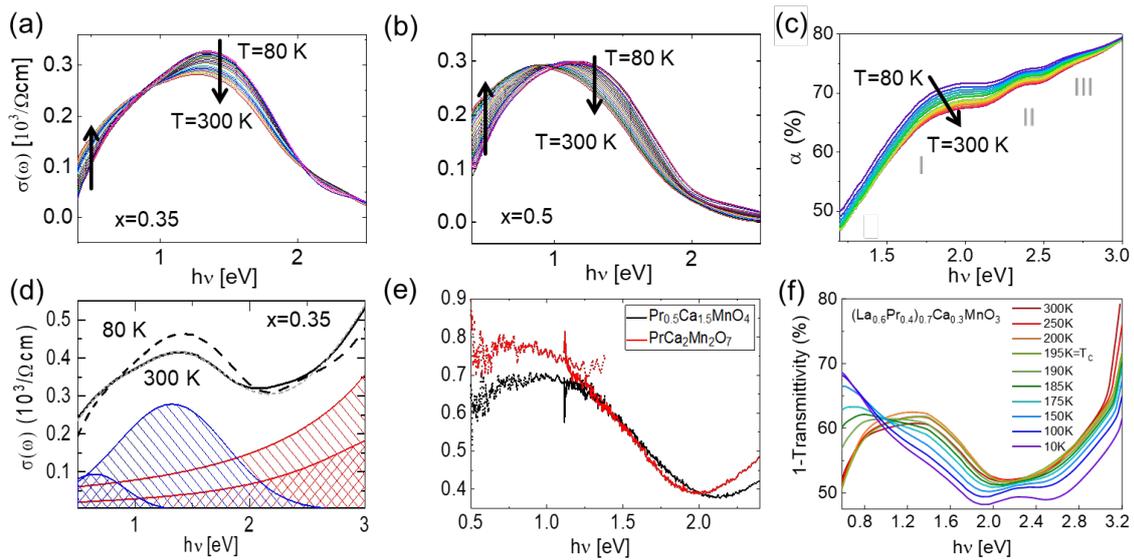

Figure 5.3: Small polaron optical conductivity due to the JT-type of small polaron optical transition for several manganites. The optical conductivity is derived from optical absorption spectroscopy at epitaxial thin films: $Pr_{1-x}Ca_xMnO_3$ for charge ordered ground states at x=0.35 (a) and x=0.5 (b) (ref. [122]) as well as the orbital ordered ground state at x=0.1 (c) (ref. [22]). In (a) and (b) the optical conductivity related to the charge transfer transitions have been subtracted after fitting, shown in (d). Small polaron optical conductivity in the RP $Pr_{0.5}Ca_{1.5}MnO_4$ (n=1) (ref. [42]) and $PrCa_2Mn_2O_7$ (n=2) (e). e) Change from small polaron to Drude-like optical absorption upon crossing the insulator metal transition in $(La_{0.6}Pr_{0.4})_{0.7}Ca_{0.3}MnO_3$ (Ref. [123]).

## 6. Relaxation and annihilation of QPs after optical excitation

Fast optical excitation of matter holds the promise to access its non-equilibrium properties and to elucidate energy transfer between different DOF like electrons, lattice, and spin. Moreover, in simple metallic systems, where a separation of these DOF is usually appropriate, this approach has allowed to obtain values for the strength of electron-phonon and spin-lattice coupling by using simple thermal models with heat baths and coupling constants given by a set of differential equations[131]. Such approaches are as well useful in the case of correlated oxides, as has been shown by numerous studies, e.g., on the CMR manganite $La_{0.7}Ca_{0.3}MnO_3$ (LCMO)[132],[133],[134]. In particular, for this system, which shows an FMM to PM insulator transition, the appearance of an unusual, long-lived transient reflectivity signal after a femtosecond optical stimulus (typically at 1.55 eV photon energy) upon entering the metallic state has been detected in many studies[135],[136]. The physical origin of such slow decay times has been attributed to different effects, such as long-living localized in-pseudogap states[136], changes in heat transport[133] and long magnon lifetimes[137], and are not yet clarified in systems with a metallic ground state.

The presence of many-body effects in correlated oxides allows for the creation of novel non-equilibrium states via fast photon excitations that are not achievable through thermodynamic pathways. This concept of optical phase control, being an active field of research in correlated oxides for nearly three decades, resulted in numerous examples highlighting the creation of photoinduced states[138],[139],[140] also studying early time dynamics in LCMO and LPCMO systems. One particular question not intensively studied so far in the literature is how to engineer the lifetime of such emergent transient states. Here, one recent study from Abreu and coworkers[141] shows that in nickelates the type of phase transition (1$^{st}$ vs. 2$^{nd}$ order) seems to strongly influence the recovery of the system from a photoinduced metallic into the insulating ground state. A similar analogy could be valid when comparing the manganites LCMO and LPCMO.

**6.1 Slow thermalization of optical excitations due to magnetic coupling**

The study on the prototypical CMR manganite $(La_{0.6}Pr_{0.4})_{0.7}Ca_{0.3}MnO_3$ (LPCMO) using a femtosecond infrared pump (1.2 eV) laser and probing reflectivity at 2.4 eV revealed that indeed a sudden increase in the recovery time of the transient reflectivity signal appears at the phase transition[123] (see Fig. 6.1) in agreement with earlier reports on LCMO[133],[135],[136]. Moreover, this signature can be controlled by applying a moderate magnetic field at the phase transition temperature (see Fig. 6.1 b)), as it is possible to switch between fast and slow ground state recovery. Modelling the system dynamics after optical excitation by using a finite difference time domain method and solving a three-temperature-model[142] (3TM) by including thermal diffusion, revealed that heat transport cannot be responsible for the increase of recovery time at $T_C$. Moreover, direct measurements of the thermal conductivity of LPCMO thin films (see sec. 4.1) demonstrated that thermal conductivity increases upon entering the FM phase and thus would lead to a faster decay. Instead, a drastic increase in the transient reflectivity recovery time at the MI transition is due to the divergence of the spin specific heat at $T_C$. Thereby, the long lifetime of the transient reflectivity signal entering the FM phase is caused by transferring and storing of energy from the hot lattice into the spin system due to the appearance of lattice-spin coupling. Close to the phase transition this storage is particularly effective due to the increase in the spin specific heat[135]. The recovery of the spin ordered state takes on the order of nanoseconds, as corroborated by measurements of magnetization dynamics. Consequently, many-body physics of polaronic QPs in this material[67] plays only the role of controlling the ground state of the material which can be modified by a magnetic field as seen in the CMR effect. The actual non-equilibrium relaxation dynamics after optical excitation is dominated by thermal effects.

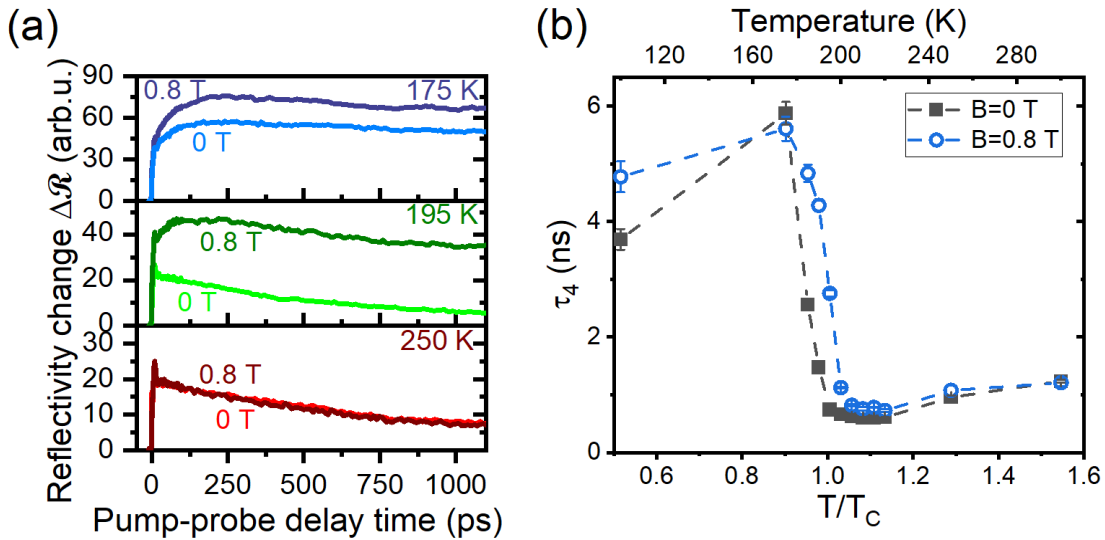

Figure 6.1: a) Transient reflectivity change without and with applied magnetic field for temperatures within the FM metallic phase (blue hues), at the phase transition temperature (green hues) and in the PM insulating phase (red hues). b) Change in relaxation time back to the ground state with temperature without applied field (black squares) and for an applied field of 0.8 T (blue open circles). Dashed black and blue lines are a guide to the eye. Figures adapted from Seick et al[123] in accordance with the *Creative Commons Attribution 4.0* (CC BY 4.0) license.

**6.2 Beyond thermal effects: Long-living optical excited state due to phase separation**

Inspecting the phase transition region in LPCMO reveals that the FM state is closely interconnected with the presence of correlated JT polarons leading to a phase separated state[13],[67]. In fact, as discussed in detail in ref. [67], at the 1st order phase transition temperature, LPCMO is expected to consist of FM nanodomains AFM coupled by correlated JT polarons located at the domain walls. This enforces a polaronic insulating state in the phase transition region with metallic FM domains and insulating polaronic phase at the domain walls. The strong CMR effect in LPCMO is thus connected to the break-up of this AFM coupling and the activation of electronic hopping between FM ordered nanodomains (see Fig. 6.2a). The strong interconnection of spin and polaron DOF leading to significant spatial correlation of polarons in LPCMO at the domain walls, hints at the potential to create long-living non-equilibrium states in this material by conversion of polaronic QP into weakly interacting electrons using laser light. In particular, annihilation of polarons should potentially lead to a nonthermal metallization of the system by liberating carriers, as also demonstrated in strained LCMO films[143],[144].

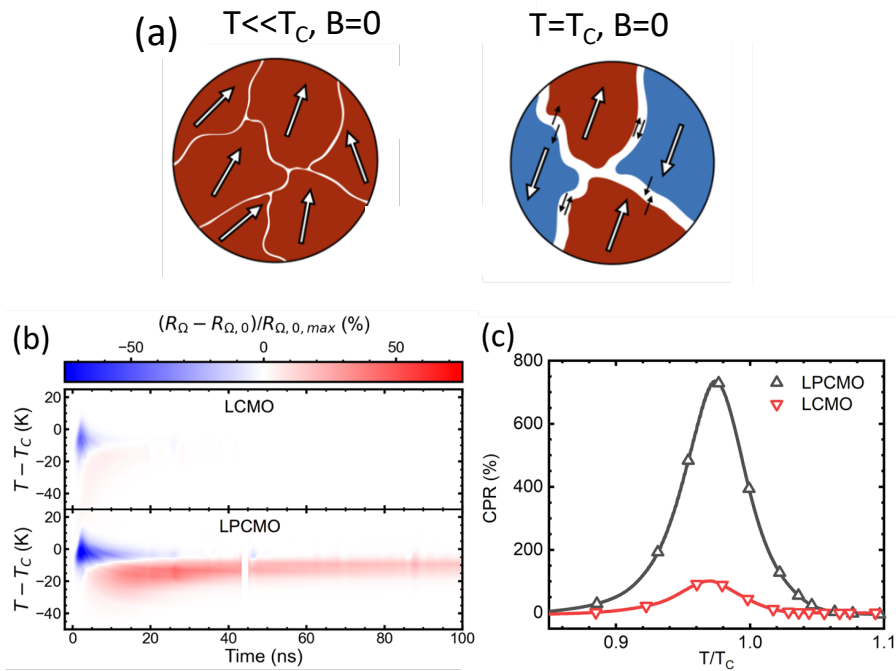

Figure 6.2: (a) Sketch of proposed magnetization state below and at $T_C$ for different applied fields. (b) Nonthermal resistance change for LCMO (top) and LPCMO (bottom) after excitation with a 1.25 ns 515 nm laser pulse. (c) Colossal photoresistance at 2 ns for F=1.5 mJ/cm$^2$ incident fluence for LPCMO and LCMO. Figures adapted from ref. [145] in accordance with the *Creative Commons Attribution 4.0* (CC BY 4.0) license.

In principle, the need to reform a certain ordering to recreate the ground state of the system in LPCMO should allow for the creation of a long-living metallic state. To elucidate the effects of polaronic correlations and spin order resulting in the observed electronic phase separation the response of optimally doped LCMO and LPCMO films to light stimulation was compared by direct measurements of the transient resistance change on the nanosecond timescale[145]. The resulting data, corrected for thermally induced resistance changes, reveals nonthermal metallization effects in the phase transition region for both LCMO and LPCMO, as depicted in Fig. 6.2 b). Crucially however, in optimally doped LCMO (x=0.3), where electron-phonon interaction and polaronic correlations are weaker but double-exchange interaction is stronger than in the LPCMO, and phase separation is not pronounced[146], these metallization effects are weaker than in LPCMO and exist mostly only on the timescale of the laser pulse. In contrast, for LPCMO, strong nonthermal metallization effects survive up to about 30 ns, or an order of magnitude longer than the laser excitation. A maximum "colossal photoresistance" (CPR) of 95% and 750% can be defined for LCMO and LPCMO, respectively, similar to the CMR values observed at a field B=1 T for these samples as seen in Fig. 6.2c).

The reason for the difference in the transient metallic state lifetime of LCMO and LPCMO can be directly connected to the polaronic correlations and sample microstructure. In LCMO, depending on the doping, polarons can form a weakly interacting polaron gas or liquid as well as a charge ordered phase above half doping. Laser excitation can break up these polarons, initially yielding the liberation of electrons and weakly metallize the sample during the presence of the laser pulse. However, polarons in LCMO reform quickly after annihilation and the disordered polaronic state reappears after the laser pulse is gone. In contrast, in LPCMO the annihilation of the correlated polarons in the domain walls destroys the AFM coupling and leads to the establishment of a more FM state, which is the ground state of the system without CPs. In this state, electronic hopping across the domain walls is possible (double-exchange), leading to the observed strong nonthermal decrease in the resistance. The polarons again reappear after the laser excitation, however, at random positions in the sample. To reestablish the initial, strongly insulating state, these polarons need to reform the correlated polaronic state within the FM domain walls and enforce the AFM coupling of FM nanodomains. The reestablishment of this correlated state represents a crucial bottleneck, which is absent in LCMO, and can serve as a blueprint for creating long-living nonthermal "hidden" states in CMR manganites.

## 7. Stabilization and transformation of excited QPs into photovoltage

Energy conversion in photovoltaics, thermoelectricity or (photo-)electrochemistry relies on materials where QP excitations e.g. stimulated by light are transformed into other forms of energy. The fundamental principle of energy conservation requires that the energy uptake of QPs by absorption of a photon quanta $h\nu$ is fully converted into thermal, electrical or chemical energy. Here, first the photon energy is converted into the chemical energy of excited QPs $\Delta\mu$ which can be identified as the splitting of the quasi-Fermi levels of the generated electron and hole. The gained $\Delta\mu$ can be further converted into a photovoltage U or Gibbs free energy of generated chemical species $\Delta g$[147] which typically requires interfaces of the absorber material with another solid or liquid.

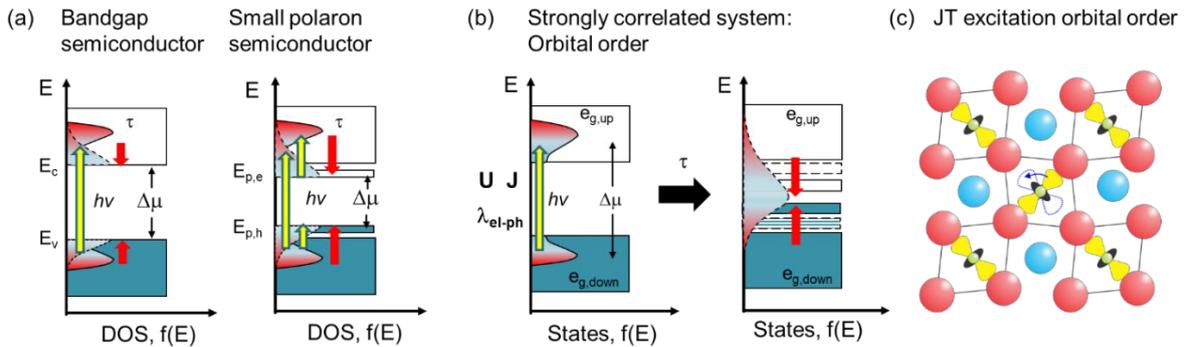

Figure 7.1: Band structure and stabilization of quasiparticle excitations by photons in materials with different correlations. (a) Electron-hole pair excitation in a conventional semiconductor with rigid density of states (DOS) bands and relaxation/thermalization described within single particle distribution functions f(E). Polaron quasiparticles due to introduction of electron-phonon interaction lead to band renormalization with polaron states of electrons and holes $E_{p,e}$ and $E_{p,h}$, respectively. Stabilization of electron-hole pair excitations requires to reduce non-radiative recombination channels, e.g. by defects or polaron localization. (b) Quasiparticle excitations in correlated systems for the example of an orbital order. Here band structure is governed by coulomb U, exchange J and electron-phonon interactions $\lambda_{el-ph}$. Upper $e_g$ states which are populated by optical excitations can be stabilized in ordered states of cooperative Jahn-Teller distortions, such as in orbital ordered states. (c) Scheme of the ordering of $e_g$ orbitals in orbital ordered states.

The well-established theoretical models for such transformation processes mostly rely on the description of electron states within the rigid energy bands, i.e. they do not couple to the structural dynamics and do not change upon excitations. This approximation works very well for conventional semiconductors (Fig. 7.1a)), where excited QPs relax and then thermalize towards band edges before they are converted into other forms of energy. In this framework, interactions such as electron-phonon or Coulomb can be taken into account by band renormalization, where polarons or excitons represent dressed electrons, i.e. QPs with modified energy and effective mass. Even in direct bandgap semiconductors, where radiative recombination for electrons and holes with equal momentum states is easily possible, the lifetime of optical QP excitations can be controlled by non-radiative recombination, i.e. Auger recombination[148] and Shockley-Read-Hall recombination at defects[149],[150]. In oxide semiconductors with small polaron band renormalization such as $TiO_2$[151] or $Fe_2O_3$[152], the electron-hole polaron recombination is largely controlled by the degree of localization of small polarons. The upper limit of transformable chemical energy in such systems is thus limited by the semiconductor bandgap that limits the splitting of the quasi-Fermi levels of electrons and holes as introduced by Shockley and Queisser[153]. Harvesting of hot carriers before they relax to the band edges into other forms of energy is thus a general strategy to achieve higher efficiencies in photovoltaics or photochemistry, since relaxation and thermalization losses of QP excitations can be reduced[154]. This requires a long enough lifetime of excited QPs to be extracted out of the absorber.

### 7.1 Stabilization mechanism of hot polaron QPs in ordered states

Stabilization of hot QPs was theoretically predicted in Mott insulators based on Coulomb repulsion[155],[156] and experimentally realized in a $LaVO_3/TiO_2$ junction[157]. Here we will focus again on the 3d-transition-metal perovskites with JT polarons and ordering of such polarons in orbital/charge ordered ground states with different magnetic structures. As introduced in section 5 (see Figure 5.1), two split bands of 3d-$e_g$ states can evolve by the JT-induced lifting of the degeneracy of the $e_g$ states accompanied by a deformation of the Mn-$O_6$ octahedra. Here, an optical excitation induces an occupation of the upper $e_g$ state, which does not match the JT distortion of the involved octahedral[121]. Consequently, a coupled reorganization process of both the electronic levels as well as the local lattice sets in, where the excited upper $e_g$ state relaxes down together with a change of the JT distortion to reach the lower $e_g$ state level. As indicated in the scheme of Figure 5.1, in the case of ordered JT distortions such as in an orbital ordered state of a manganite[22], the relaxation dynamics will be strongly affected by the ordering. Such a coupled electron and structural dynamics thus offers new pathways of stabilization of excited QPs e.g. by slowing down the structural relaxation of hot polaron states by strong correlations.

Figure 7.2 shows the transient optical dynamics of hot polarons after optical pumping at the maximum weight of the optical small polaron transition at hv = 1.55 in a $Pr_{0.65}Ca_{0.35}MnO_3$ manganite thin film[121]. After fs excitation, there is a fast relaxation process $t_{fast} \sim 900$ fs to a $\delta$-type of potential energy surface, where radiative recombination with the hole state is forbidden and thus a long living hot state can be stabilized with a lifetime of $t_{slow} \sim 1$ ns. The hot state is dynamic and initially shows a cooperative type of coupled electron-phonon dynamics with a frequency of about 2 THz (Figure 7.2 b)). In depth analysis shows that there are three main modes at around 1 THz (4.2 meV), at 2.2 THz (9 meV) and 3.9 THz (16 meV) which all evolve only below the charge/orbital ordering temperature $T_{CO} \approx 240$ K. The oscillation in $\Delta OD$ with $t_{coh} \approx 40$ ps is due to a pump-induced thermo-elastic wave that travels forth and back across the film. Similar processes are observed by transient optical pump-probe studies also in other orbital or charge/orbital ordered manganites[22] and thus establish the presence of a long living hot polaron state in the ordered ground states of the phase diagram.

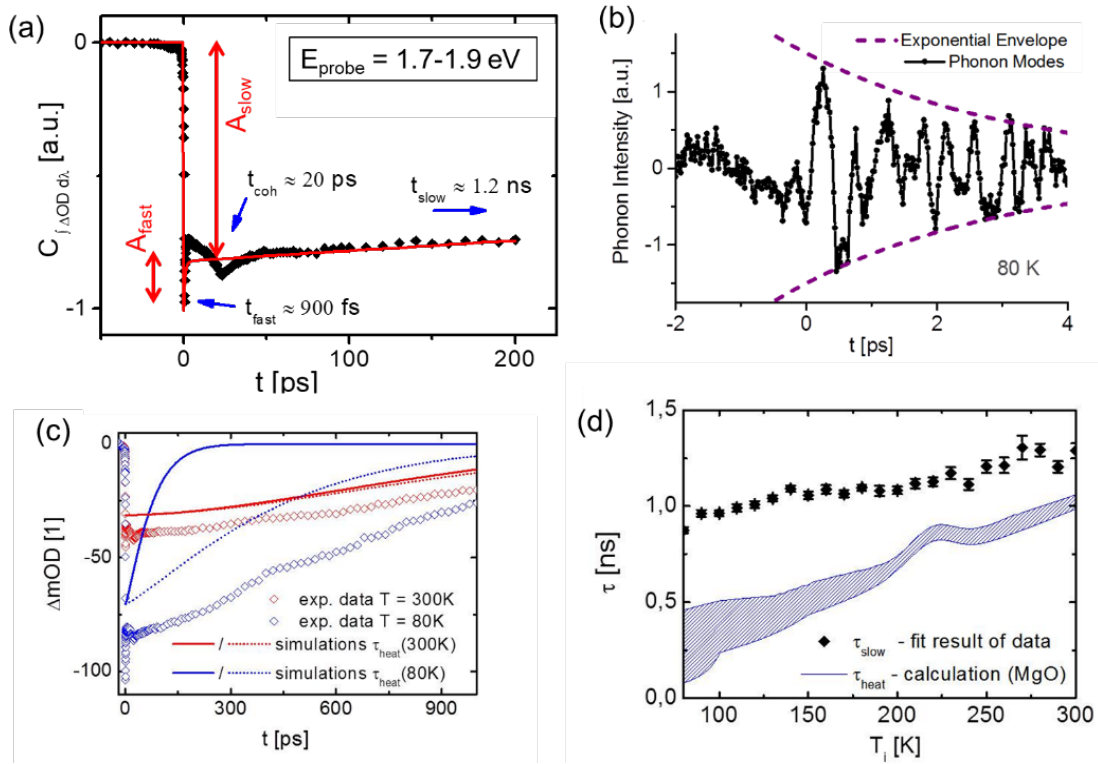

Figure 7.2: Hot polaron quasiparticle with ns lifetime in a charge and orbital ordered ground state of a manganite. (a) Time-resolved transient change of the optical density $\Delta OD$ at $E_{probe}$=1.7-1.9 eV of an epitaxial $Pr_{0.65}Ca_{0.35}MnO_3$ thin film after ultrafast photoexcitation at hv = 1.55 eV. Black: data. Red line: multiexponential fit. (b) Excited state polaron dynamics at T = 80 K with frequency of around 2 THz as revealed by subtraction of the exponential fit from the measured optical density. (c) Time decay of the slow fraction of $\Delta OD$ compared to the simulated effect

of $\Delta OD$ due to heating and heat transfer across the substrate. (d) Temperature-dependent live time of the hot polaron compared to the thermalization time. Redrawn from ref. [121].

### 7.2 Hot polaron photovoltaics in junctions with strong correlations

Stabilization of hot QPs in an absorber material is one of the two prerequisites for the establishment of hot carrier photovoltaic or photochemical energy conversion. The second one is a suitable interface which enables charge separation of electrons and holes to convert chemical into electric potential. The majority of experimental research on correlated electron photovoltaics has been performed on junctions of a manganite film to n-doped STO[158],[159],[160],[161], where excess electron polarons can be transferred from the manganite to the titanite, whereas hole carriers are blocked.

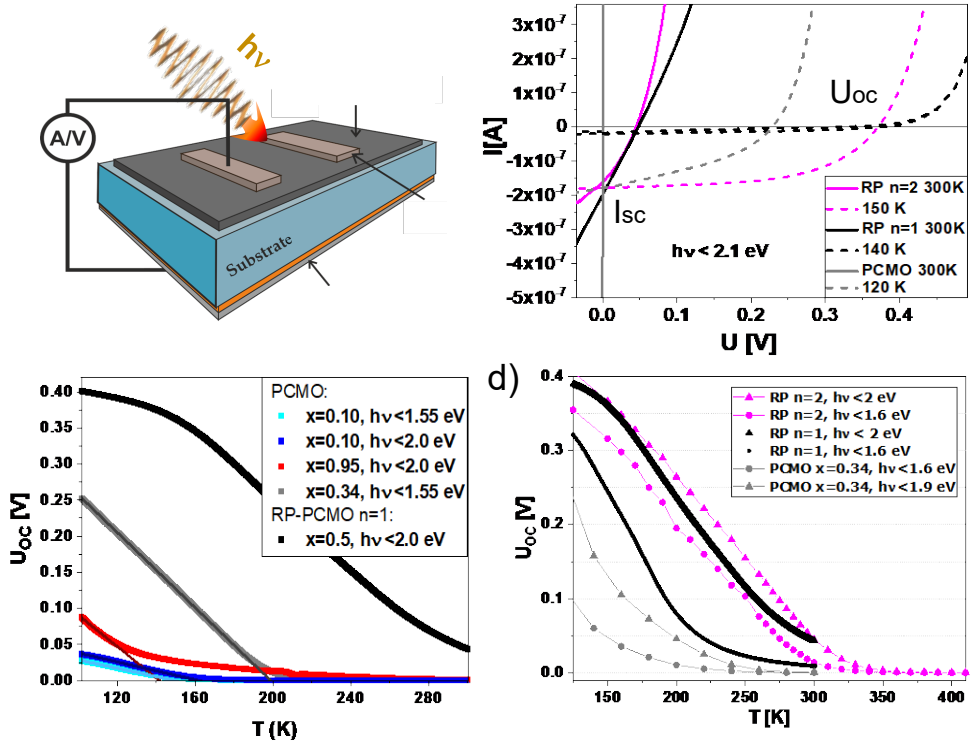

Figure 7.3: Photovoltaic effect in manganite-titanite junctions. (a) Scheme of the geometry of the junctions with front and back contacts. (b) Rectifying current-voltage characteristics of the junctions with different RP variants (n=1, 2 and ∞) of the $(CaO)(PrCaMnO_3)_n$ system used as absorbers under illumination at the IR spectral range (hv ≤ 2 eV) at T = 300 K. (c) Temperature-dependent open circuit voltage $U_{oc}(T)$ for the three-dimensional n=∞ system at various Ca-doping x=0.1, 0.34 and 0.95, where the onset of the photovoltaic effect is correlated to the ordering phase transition temperatures (ref. [42]). (d) Comparison of the temperature-dependent open circuit voltage $U_{oc}(T)$ for the n=1, 2 and ∞ systems in the IR.

Figure 7.3 shows photovoltaic properties of manganite-titanite junctions consisting of 3D perovskite $Pr_{1-x}Ca_xMnO_3$ with x=0.1, x=0.35 and x=0.95 as well as of the layered n=1 and n=2

Ruddlesden-Popper variants (PrO)(Pr$_{1-x}$Ca$_x$MnO$_3$)$_n$) at half doping (x=0.5). Here, the temperature-dependent onset of the open circuit voltage U$_{oc}$ is ultimately connected to the onset of the phase transitions to orbital ordering (x=0.1), charge/orbital ordering (x=0.35). In contrast, in the x=0.95 system with the dominant Mn$^{4+}$ state and without JT effect, photovoltaics only occurs at the charge transfer transition[162], thus the hot polaron effect is absent. While the effect was found over many years only at low temperatures in the ordered phases of the 3D manganites, it could be shifted above room temperature using the 2D layered RP films with n=1 (ref. [42]) and n=2 as absorber materials with higher ordering temperatures of T$_{CO}$=320 K (n=1) and T$_{CO}$=380 K (n=2), respectively. Here, open circuit voltages of up to 240 mV at room temperature and up to 500 mV at low temperature could be achieved.

Figure 7.4 a) shows a band scheme for a hot polaron manganite titanite junction, where charge separation of the excess carrier QPs occurs due to blocking of polaron hole and transfer of polaron electron states above a certain energy threshold $\Delta E$. However, the transfer may be limited by the low orbital overlap of upper Mn 3d eg and Ti 3d t2g states due to different orientations of the orbitals. Figure 7.4 b) shows a specific example for a PrMnO$_3$ junction. Charge separation requires the transport of hot polaron quasiparticles from the absorber to the interface.

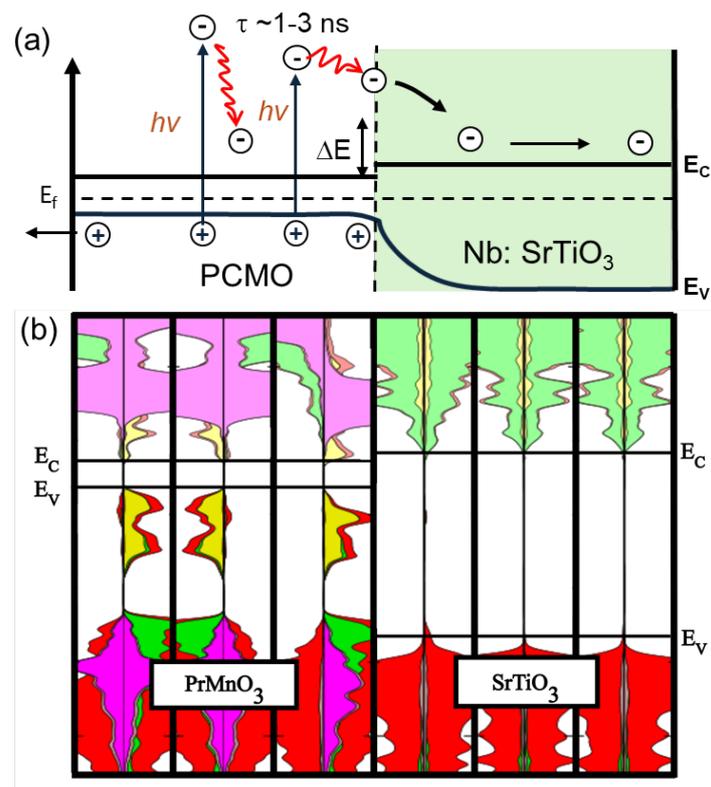

Figure 7.4: Band scheme of the charge separating interface of manganite-titanite junctions. (a) Charge separation after optical excitation of hot polarons due to the selective interface, where hole polarons from the manganite are blocked and electron polarons can be transferred above a certain energy threshold $\Delta E$. (b) Ab-initio calculation of the density of states of a PrMnO$_3$ manganite – SrTiO$_3$ titanite junction as a function of position across the interface. The density of states projected onto oxygen is shown in red, the one projected onto the transition metal ions (Mn,Ti) are shown in yellow for 3d-e$_g$ states and green for 3d-t$_{2g}$ states. The Pr-f states are magenta. Light colors mark the unoccupied, dark color the occupied states.

Due to the small mobility of the polaron quasiparticles, the diffusion length is quite small, i.e. 2-3 nm at room temperature and around 10 nm at low temperatures[163],[164]. The small ratio between the diffusion length $l_d \approx$ 3-5 nm and absorption length $l_a \approx$ 70 nm of l$_d$/l$_a \approx$0.04-0.07 in manganites limits the achievable photocurrents and thus the hot polaron photovoltaic conversion efficiency. Remarkably, based on the presence of colossal magneto- and electroresistance effects in the manganite absorber system, the photovoltaic effect can be strongly influenced by magnetic[165],[166] and/or electric field[167].

### 7.3 Spectral and energy density dependence of hot polaron photovoltaics

The fact that hot polaron quasiparticles represent the essential vehicles for optical to electric energy conversion in such manganite junctions is most clearly visible from the spectral and power density dependent photovoltaic response shown in Figure 7.5. For a conventional bandgap semiconductor, where under stationary conditions all excited electron and hole quasiparticles have relaxed to the band edges before being extracted across interfaces, Shockley-Queisser theory yields for the temperature dependence of $U_{oc}(T)$ [153],[168]:

$$qU_{oc} = \Delta E - k_B T \ln\left(\frac{N_C N_V}{n_e n_h}\right) \qquad (1)$$

Here the spectral and power density dependent response enters via the change of electron density $n_e(h\nu, p)$ and hole density $n_h(h\nu, p)$. Since photon absorption generates an excess carrier density by generation of electron-hole pairs, these quantities both depend on power density of the absorbed photons and on the photon energy, e.g. due to a spectral dependent absorption coefficient. The energy barrier $\Delta E$ at the junction as well as the effective densities of states at the band edges of the valence and conduction bands $N_v$ and $N_c$, respectively, do not change upon illumination. The temperature dependence of $N_v$ and $N_c$ generates only a small correction to the linear $U_{oc}(T)$ behavior. Thus, the close-to linear temperature dependence of $U_{oc}(T)$ is a fingerprint of rigid states, where excited carriers are relaxed to band edges before they are extracted at the junction's interface. For weak generation, the concentration of photoinduced excess carriers $\Delta n_e$ and $\Delta n_h$ is directly proportional to the spectral and power dependent generation rate $G(T,p,E_{ph})$, thus, eq. (1) directly leads to a logarithmic illumination power density dependence of $U_{oc}(T,p)$ and a step like spectral $U_{oc}(T,E_{ph})$ dependence, where $U_{oc}$ is independent of photon energy for $E_{ph} > E_g$.

In contrast, the hot polaron junctions in correlated manganites show a scaling law type of behavior[169]

$$U_{oc}(T, p, E_{ph}) = U_o(h\nu, p) \left(1 - \frac{T}{T_0(h\nu,p)}\right)^\beta \quad (2)$$

with a critical exponent β and spectral and power density dependent scaling parameters $U_o(h\nu, p)$ and $T_0(h\nu, p)$, respectively. The scaling law analysis of measured spectral and power density dependent open circuit voltages for low doped and half-doped manganite-titanite junctions is shown in Figure 7.5 c)-h). One yields critical exponents of β = 3.5 for both junctions demonstrating that the close-to linear behavior of eq. (1) is not found experimentally.

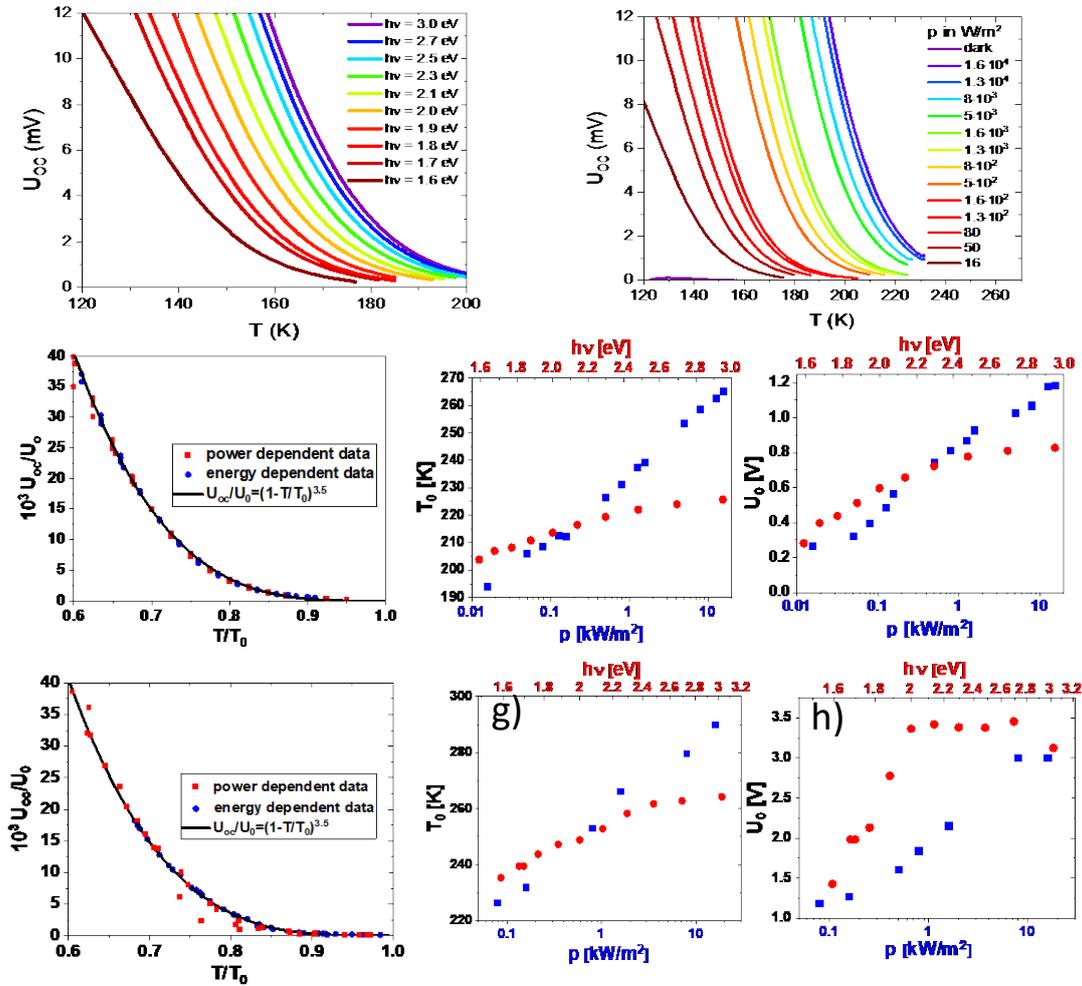

Figure 7.5: Spectral and power density dependent photovoltaic response of hot polaron $Pr_{1-x}Ca_xMnO_3$-Nb:SrTiO$_3$ junctions. (a) Spectral $h\nu$ and (b) power density $p$ dependent open circuit voltage $U_{oc}(T)$ for x=0.1. (c) Scaling behavior of the temperature dependent open circuit voltage for x=0.1, normalized to the parameters $T_0(h\nu,p)$ and $U_0(h\nu,p)$ which are shown in (d) and (e), respectively. The black line corresponds to a critical exponent of y=3.5 and normalized experimental $U_{oc}(T)$ at different power densities (green) and photon energies (blue) are shown as symbols (ref. [169]); (f-h) Same as (c-e) for x=0.5.

That both the onset temperature $T_0(h\nu)$ and $U_o(h\nu)$ scale linearly with photon energy hν in the range of JT polaron absorption and saturates in the range of charge transfer transitions reflects the increased splitting of the chemical potential of hot electron-hole polaron pairs with photon energy. This is a distinguishing feature of the hot carrier junctions. Furthermore, both $T_0(p)$ and $U_o(p)$ scale logarithmically with photon power density p, reflecting how the non-equilibrium Fermi-Dirac statistics depends on the concentration of excess carriers. This dependence naturally exists both in conventional semiconductors as well as in hot carrier junctions.

In contrast to conventional semiconductor junctions, where the onset temperature of $U_{oc}(T)$ is controlled by the thermally activated generation current across the rigid junction barrier $\Delta E$, in the hot polaron systems the onset is strongly affected by the phase transitions to the charge/orbital ordered states since ordering is a prerequisite for long lifetimes of hot polaron QPs. Indeed, the range of values for the non-equilibrium scaling parameter $T_0(h\nu, p)$ covers the experimentally observed equilibrium transition temperatures of the phase transition $T_{oo}$≈250 K (x=0.1) and $T_{co}$≈240 K (x=0.5), however, it is not the same. Photovoltaic and photochemical energy conversion based on correlated electron absorber materials thus show a huge degree of tunability of the conversion process, e.g. by photon energy and power density. Furthermore, photovoltaics can be tuned by phase transitions and different types of order, since they change types of quasiparticles and their lifetimes.

The finding of a scaling law for the non-equilibrium property $U_{oc}(T, p, E_{ph})$ due to the non-equilibrium response of polaron quasiparticle excitations across ordering phase transitions is the first experimental support to the general question, whether scaling laws exist in non-equilibrium phase transitions. Until now, the existence of the underlying critical behavior was discussed theoretically[170] for the special class of so-called "absorbing" phase transitions. Here, a non-equilibrium phase transition between a non-fluctuating "absorbing" state (e.g. the ground state) and a non-equilibrium state occurs via critical fluctuations above a critical threshold parameter[170],[171]. Attempts to find critical exponents in such systems according to different universality classes are still at the infancy[172].

## 8. Conclusions

Elementary steps of energy conversion – excitation, relaxation and transformation – have been studied in well characterized thin films and superlattices of 3D perovskite manganites and 2D layered Ruddlesden-Popper variants with the goal to develop a fundamental understanding of energy conversion in strongly correlated electron materials. Manganites are particularly suitable as model system, since their ground state can be easily tuned by A-site substitution, allowing to influence the subtle balance of magnetic, Coulomb and electron-phonon interactions which give rise to their rich phase diagrams. In addition, heterovalent A-site doping enables to modify the QP type and density, ranging from large band type polarons in ferromagnetic metallic to small rather localized Jahn-Teller type polarons in insulating or semiconducting phases which can have different types of magnetic order. The type of lattice distortion, electronic structure and ordering of 3d type of Jahn-Teller polarons can be influenced by the octahedral tilting as well as by the degree of hybridization with O2p states, forming so called Zener polarons. Furthermore, these polarons, depending on their density and type, can exhibit various types of ordering, including polaron crystals (charge and orbital order), liquid and disordered, even glass-like structures. Since the QPs are the main vehicle for energy conversion, their tunability upon doping, temperature, excitation and via interfaces requires the introduction of qualitatively new theoretical concepts to describe energy conversion beyond single particle and rigid band approximations.

The low-energy excitations and relaxations of polaron QPs have been studied by measurements of *local friction* and *thermal conductivity* in 3D manganite films of LSMO, LCMO, LPCMO and PCMO. It was demonstrated that both friction and thermal transport is strongly affected by phase transitions which in turn are governed by modification of polaron QPs behavior, e.g., from large to small polarons at the phase transition from ferromagnetic metallic to paramagnetic insulating state in LSMO. Moreover, the octahedral tilt/rotation or Mn-O-Mn bond angle $\varphi_{OOR}$ in manganites has been shown to be a generic structural control parameter for tuning the electronic bandwidth which in turn determines the type of polaron QPs and controls thermal conductivity. The observed strong dependence of thermal transport and friction on the interface/surface properties has been rationalized by means of the corresponding changes of $\varphi_{OOR}$ at the interfaces. These interface-related changes of $\varphi_{OOR}$ angle underlie the correlations between the condensation of polaron QPs at the interfaces and formation of layers of localized polarons with charge/orbital ordered AFM ground state. The polaron ordering at the interface can be melted by applied electric field, resulting in a low-resistance metallic state and suppression of local friction in LCMO film at room temperature.

The "high-energy" optical excitations and relaxations of polaron QPs have been studied in LCMO, LPCMO and PCMO films by time-resolved pump-probe optical and electric measurements. A strong temperature and magnetic field dependence of the relaxation time of optical reflectivity has been observed in the vicinity of phase transition yielding an extreme slowing down of the relaxation of excited QPs. The reason for this behavior is the divergence of specific heat of the spin system at the phase transition, allowing to absorb a significant part of the energy of excitations and thus to damp directly the thermal relaxation. Our results demonstrate a fundamental role of spin-phonon coupling in setting the long time scale (up to 100 ns) of relaxation of QPs after optical excitations. Along with a thermal optical and electric response, non-thermal or photoinduced effects have been observed in LPCMO films and assigned to the melting/condensation of correlated JT polarons at the magnetic domain walls of the phase separated LPCMO. After photoexcitation of LPCMO films, a metastable metallization has been detected and found to be especially pronounced and long-living in the vicinity of the phase transition. Thus, formation and annihilation of phase separated states represents one general underlying mechanism for long lifetimes of QP excitations, since relaxation is slowed down by their diffusion and transformation through formation of phase boundaries.

A second mechanism for slowing down hot QP excitations was found in charge and orbital ordered phases, where a stabilization occurs via ordering-induced cooperative response to the hot-polaron-related orbital disorder. The occupation of $e_g$ states by hot QPs with a non-consistent Jahn-Teller distortion creates cooperative electron-phonon dynamics in the THz range and its relaxation towards electron-hole recombination is damped most probably due to the required reorganization of the local orbital microstructure. Such effects are observed in the orbital and charge ordered phase of PCMO and layered Ruddlesden-Popper PCMO (n=1,2) films. Combining such hole doped films with a n-doped substrate (Nb:STO), the chemical energy of excited hot electron-polaron pairs can be harvested into a photovoltage by charge separation at the PCMO/Nb:STO interface. The illumination and temperature dependent photovoltaic response of such junctions fundamentally differs from that of conventional semiconductor junctions due to the change of QPs by the presence of the phase transitions as well as due to the strong dependence of electronic band structure on excitation and relaxation of small polarons.

The fundamental role of Jahn-Teller physics for the understanding of ground state and polaron QPs in manganites could be generalized onto other types of correlated electron systems, such as nickelates, cuprates and ruthenates. Indeed, Ni, Cu and Ru, depending on their oxidation state, can exhibit weak up to strong Jahn-Teller effects. In nickelates, the Jahn Teller polaron effect is discussed to be the origin of small-polaron-hopping-related electric transport in $SmNiO_3$[173] and the origin of charge ordering of stripe type in $La_{2-x}Sr_xNiO_4$[174]. The research on Jahn-Teller polaronic distortion in cuprates was the starting point of the detection of high-temperature superconductivity[11]. It is considered to play a major role in the formation of a condensate of cooper pairs[175]. The JT effect has recently been considered as a possible factor in the superconducting states of ruthenates[130]. Understanding the rich phenomena of tuning and transformation of polaronic QP properties and their excitations in the different ground states by rather subtle changes of interactions opens up a new window in studies of ordered low temperature phases and can establish a new paradigm of tunable excitations, energy transport and energy transformations in advanced energy materials.

## Acknowledgements

This research is funded by the Deutsche Forschungsgemeinschaft (DFG, German Research Foundation) 217133147/SFB 1073, projects A01, A02, B02 and B03 and Z02.